\documentclass[aps,preprint,tightenlines,floatfix,superscriptaddress,
nofootinbib,showpacs]{revtex4}
\usepackage{graphicx}
\usepackage{dcolumn}
\usepackage{amsfonts}
\usepackage{multirow}
\usepackage{longtable}
\usepackage{amsmath}
\usepackage{subfigure}
\def\beq{\begin{equation}}
\def\eeq{\end{equation}}
\def\bea{\begin{eqnarray}}
\def\eea{\end{eqnarray}}
\def\nn{\nonumber}

\def\roughly#1{\mathrel{\raise.3ex\hbox
{$#1$\kern-.75em\lower1ex\hbox{$\sim$}}}}

\def\gsim{\roughly>}

\def\tbar{{\overline{t}}}
\def\bbar{{\overline{b}}}
\def\nubar{{\overline{\nu}}}
\def\tbbc{t \to b \bbar c}
\def\ggprocess{gg\to t\tbar\to\left(b\bbar c\right)
    \left(\bbar\ell\nubar\right)}
\def\ggprocessenu{gg\to t\tbar\to\left(b\bbar c\right)
    \left(\bbar e^- {\overline{\nu}}_e \right)}
\def\tm#1{\texttt{TM-{#1}}}
\def\ex#1{\texttt{EX-{#1}}}
\def\Ahat{\hat A_i^{\sigma}}
\def\madg{\mbox{MadGraph~5} }
\begin{document}
\bibliographystyle{apsrev}

\preprint{\vbox {\hbox{UdeM-GPP-TH-14-234}}}

\vspace*{2cm}

\title{\boldmath Detecting New Physics in Rare Top Decays at the LHC}

\def\umontreal{\affiliation{\it Physique des Particules, Universit\'e
    de Montr\'eal, \\ C.P. 6128, succ.\ centre-ville, Montr\'eal, QC,
    Canada H3C 3J7}}
\def\tayloru{\affiliation{\it Physics and Engineering Department,
    Taylor University, \\ 236 West Reade Ave., Upland, IN 46989, USA}}
\def\laplata{\affiliation{\it IFLP, CONICET -- Dpto. de F\'{\i}sica,
    Universidad Nacional de La Plata, C.C. 67, 1900 La Plata,
    Argentina}}

\umontreal
\tayloru
\laplata

\author{Pratishruti Saha}
\email{pratishruti.saha@umontreal.ca}
\umontreal

\author{Ken Kiers}
\email{knkiers@taylor.edu}
\tayloru

\author{David London}
\email{london@lps.umontreal.ca}
\umontreal

\author{Alejandro Szynkman}
\email{szynkman@fisica.unlp.edu.ar}
\laplata

\date{\today}

\begin{abstract}
In the companion paper it was shown that there are six observables in
$\ggprocess$ that can be used to reveal the presence of new physics
(NP) in $t\to b \bbar c$. In the present paper we examine the
prospects for detecting and identifying such NP at the LHC, in both
the short term and long term. To this end, we develop an algorithm for
extracting the NP parameters from measurements of the observables.  In
the short term, depending on what measurements have been made, there
are several different ways of detecting the presence of NP.  It may
even be possible to approximately determine the values of certain NP
parameters. In the long term, it is expected that all six observables
will be measured. The values of the NP parameters can then be
determined reasonably precisely from a fit to these measurements,
which will provide good information about the type of NP present in
$t\to b \bbar c$.
\end{abstract}


\maketitle

%%%%%%%%%%%%%%%%%%%%%%%%%%%%%%%%%%%%%%%%%%%%%%%%%%%%%%%%%%%%%%%%%%%%%%
\section{Introduction}
%%%%%%%%%%%%%%%%%%%%%%%%%%%%%%%%%%%%%%%%%%%%%%%%%%%%%%%%%%%%%%%%%%%%%%

Top physics provides a fertile ground for new-physics (NP) searches.
With a mass close to the electroweak scale, the top quark may well be
sensitive to interactions that do not affect other fermions.  In
Ref.~\cite{companion}, the companion paper, top decay is investigated
for the presence of NP. It is noted that, given the good agreement
between the experimental measurement of $\Gamma_t$ and its theoretical
prediction~\cite{SMtopwidth}, significant NP contributions to top
decay can only be present in decay modes that are suppressed in the
standard model (SM). One example is $\tbbc$, whose amplitude involves
the small element $V_{cb}$ ($\simeq 0.04$) of the
Cabibbo-Kobayashi-Maskawa (CKM) quark mixing matrix.
Ref.~\cite{companion} focuses on this decay at the LHC, where top
production occurs predominantly via gluon fusion: $gg \to t \tbar$.
The goal is to find observables in the channel $gg \to t \tbar$ with
$t \to b \bbar c$ and $\tbar \to \bbar \ell \bar \nu$ that can reveal
the presence of NP.

NP contributions to the decay $t \to b \bbar c$ can be parametrized in
terms of higher-dimensional operators.  If one restricts to 
dimension-6 operators, then this is realized in the form of ten
operators that span all possible Lorentz structures.
In Ref.~\cite{companion}, two types of observables are identified that
can then be used to get a handle on this NP. The first consists of
invariant mass-squared distributions involving the $\{b,c\}$,
$\{\bbar,c\}$, or $\{b,\bbar\}$ quark pairs coming from $\tbbc$.
As for the second type, we note that, in $gg \to t \tbar$, the spins
of the $t$ and $\tbar$ are correlated.  The spin-correlation
coeffcient ($\kappa_{t \tbar}$) depends only on the production
process. However, since the $t$ quark has an extremely short
lifetime, this quantity has to be inferred by measuring the angular
correlation between the decay products of the $t$ and those of the
$\tbar$. If there are NP contributions to the decay, the inferred 
value of $\kappa_{t \tbar}$ is necessarily altered from that of the 
SM. It is this feature that provides information about the NP.
Therefore, the second type of observable consists of these angular
correlations. They are taken between the $\ell^-$ coming from the
$\tbar$ decay and one of $\bbar$, $b$ or $c$ coming from the $t$ decay.

The NP operators not only change the top branching fraction of this
decay, but also modify the shapes of these distributions. It is shown
in Ref.~\cite{companion} that the NP contributions to all of the above
observables can be written in terms of certain combinations of the NP
couplings, denoted as $\Ahat$. Furthermore, the observables are found
to be practically unaffected by parton densities, etc., so that they
provide direct access to the values of these $\Ahat$'s.

Now, the observables described above involve the $\bbar$ quark coming
from the decay of the $t$.  However, there is also a $\bbar$ produced
in the $\tbar$ decay. A realistic analysis must deal with the question
of how to distinguish the two $\bbar$'s. In addition, while the focus
in Ref.~\cite{companion} was entirely on $t \tbar$ production from
gluon fusion, there is also a contribution from $q \bar q \to t \tbar$
which must be considered.

In the present paper we address these issues. In
Ref.~\cite{companion}, the analytical expressions for the observables
are compared with the results of a numerical simulation of the LHC
using \madg\cite{MG}. Here we extend our \madg\ simulations to
examine different strategies for extracting the NP parameters.  In so
doing, we include a method for distinguishing the two $\bbar$'s.  We
also take the contribution from $q \bar q \to t \tbar$ into account,
examining its effect on the aforementioned observables and their
sensitivity to the $\Ahat$'s.  In our simulations we consider numbers
of total events representative of LHC measurements in both the short
and long terms. While the long-term results obviously have smaller
errors, it is still possible in the short term to detect and partially
identify NP in $\tbbc$.

We begin in Sec.~II by summarizing the results of
Ref.~\cite{companion}.  We present the NP operators that contribute to
$\tbbc$, as well as the $\ggprocess$ observables that can reveal the
presence of the NP. In Sec.~III we develop the algorithm to extract
the NP parameters from the observables.  We discuss the
\madg\ simulations in Sec.~IV, and apply the algorithm.  Here we show
that the measurement of the observables at the LHC can lead to the
detection of the NP, and possibly even its
identification\footnote{When we refer to the
``identification of NP'', what is implied is the 
measurement of the various $\Ahat$'s and Re($X^V_{LL}$).}. We
conclude in Sec.~V.

%%%%%%%%%%%%%%%%%%%%%%%%%%%%%%%%%%%%%%%%%%%%%%%%%%%%%%%%%%%%%%%%%%%%%%
\section{New Physics in Top Decay}
%%%%%%%%%%%%%%%%%%%%%%%%%%%%%%%%%%%%%%%%%%%%%%%%%%%%%%%%%%%%%%%%%%%%%%

In this section, we summarize the main results of Ref.~\cite{companion}.

\subsection{\boldmath $\tbbc$: effective Lagrangian}

In the SM, the decay $\tbbc$ proceeds through $t\to W^+ b$, followed
by $W^+ \to \bbar c$.  The NP contributions to this can be
parameterized by the effective Lagrangian 
${\cal L}_{\mbox{\scriptsize eff}} = 
{\cal L}_{\mbox{\scriptsize eff}}^V +
{\cal L}_{\mbox{\scriptsize eff}}^S + 
{\cal L}_{\mbox{\scriptsize eff}}^T$,
with
\begin{eqnarray}
  {\cal L}_{\mbox{\scriptsize eff}}^V & = & 4\sqrt{2}G_F V_{cb}V_{tb}
       \left\{
    X_{LL}^V\,\bbar\gamma_\mu P_L t \,
       \overline{c}\gamma^\mu P_L b
   + X_{LR}^V\,\bbar\gamma_\mu P_L t \,
       \overline{c}\gamma^\mu P_R b
\right.\nonumber\\
& & \hskip2.2truecm \left.
   +~X_{RL}^V\,\bbar\gamma_\mu P_R t \,
       \overline{c}\gamma^\mu P_L b
   + X_{RR}^V\,\bbar\gamma_\mu P_R t \,
       \overline{c}\gamma^\mu P_R b
\right\}+ \mbox{h.c.}, 
\label{eq:eff1}\\
&&\nonumber\\
  {\cal L}_{\mbox{\scriptsize eff}}^S & = & 4\sqrt{2}G_F V_{cb}V_{tb}
\left\{
     X_{LL}^S\,\bbar P_L t \,\overline{c} P_L b
   + X_{LR}^S\,\bbar P_L t \,\overline{c} P_R b
\right. \nonumber\\
& & \hskip2.2truecm \left.
   +~X_{RL}^S\,\bbar P_R t \,\overline{c} P_L b
   + X_{RR}^S\,\bbar P_R t \,\overline{c} P_R b
\right\}+\mbox{h.c.,} 
\label{eq:eff2}\\
&&\nonumber\\
  {\cal L}_{\mbox{\scriptsize eff}}^T & = & 4\sqrt{2}G_F V_{cb}V_{tb}
\left\{
     X^T_{LL} \overline{b}\sigma^{\mu\nu}P_L t \,
     \overline{c}\sigma_{\mu\nu}P_L b 
\right. \nonumber\\
& & \hskip2.2truecm \left.     +~X^{T}_{RR}
\bbar\sigma^{\mu\nu}P_R t \,
     \overline{c}\sigma_{\mu\nu} P_R b
\right\}+\mbox{h.c.}
\label{eq:eff3}
\end{eqnarray}
Here the colour indices are assumed to contract in the same way as
in the SM (i.e., the fields $\bbar$ with $t$ and $\overline{c}$ with
$b$). Colour-mismatched terms, in which the indices contract
in the opposite way, may occur in certain models and can be
incorporated in a straightforward manner~\cite{kklrsw}.

The NP couplings (the $X^I_{AB}$ in the above equations) contain weak
phases, but the strong phases are negligible \cite{DatLon}. In
addition, the $X^I_{AB}$ may be assumed, quite generally, to be
$O(1)$.  The sizes of the SM and NP contributions to $\tbbc$ would
then be roughly equal. This shows that it is important to include both
the SM-NP and NP-NP interference pieces when computing the effect of
NP on a particular observable.

%%%%%%%%%%%%%%%%%%%%%%%%%%%%%%%%%%%%%%%%%%%%%%%%%%%
\subsection{\boldmath Observables in $\ggprocess$}
%%%%%%%%%%%%%%%%%%%%%%%%%%%%%%%%%%%%%%%%%%%%%%%%%%%
\label{sec:observables}

\begin{figure}[!htbp]
\begin{center}
\resizebox{4in}{!}{\includegraphics*{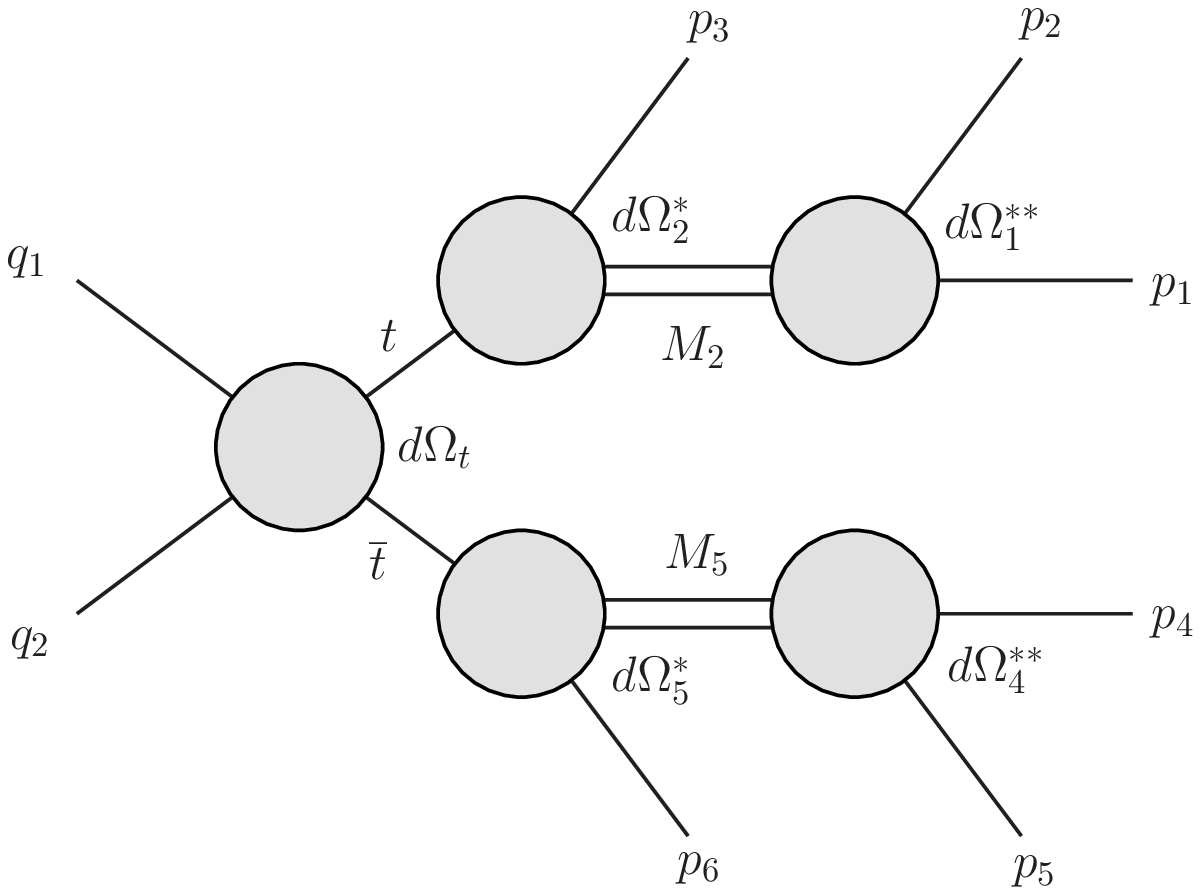}}
\caption{Kinematics for the process
  $\ggprocess$~\cite{ggprocesskinematics}.  $\Omega_1^{**}$ denotes
  the direction of $\vec{p}_1^{~**}$ in the rest frame of $M_2$,
  relative to the direction of $\vec{p}_1^{~*}+\vec{p}_2^{~*}$, where
  $M_2^2 = (p_1+p_2)^2$.  Similarly, $\Omega_2^*$ denotes the
  direction of $\left(\vec{p}_1^{~*}+\vec{p}_2^{~*}\right)$ in the $t$
  rest frame, relative to the direction of $\vec{p}_t$ in the $t\tbar$
  rest frame.  $\Omega_t$ denotes the direction of $\vec{p}_t$
  relative to $\vec{q}_1$, also in the $t\tbar$ rest frame.  The solid
  angles $\Omega_4^{**}$ and $\Omega_5^{*}$ are defined analogously to
  $\Omega_1^{**}$ and $\Omega_2^{*}$, respectively, and $M_5^2 =
  (p_4+p_5)^2$.}
\label{fig:kinematics}
\end{center}
\end{figure}

The kinematics of $\ggprocess$ is represented in
Fig.~\ref{fig:kinematics}. The six-body phase space is decomposed into
five solid angles $d\Omega_1^{**}$, $d\Omega_2^{*}$, $d\Omega_4^{**}$,
$d\Omega_5^{*}$ and $d\Omega_t$, and two invariant masses $M_2$ and
$M_5$. The $*$ and $**$ superscripts on the solid angles indicate that
these angles are defined in reference frames that are, respectively,
one and two boosts away from the $t\tbar$ rest frame. $M_2$ and $M_5$
are defined by $M_2^2 = \left(p_1+p_2\right)^2$ and $M_5^2 =
\left(p_4+p_5\right)^2$.  Note that $p_1$, $p_2$ and $p_3$ are the
momenta of the $b$, $\bbar$ and $c$ quarks in $\tbbc$, but all
permutations are allowed.  The observables use several of these
possibilities.

The differential cross section for $\ggprocess$ is computed in
Ref.~\cite{companion}. It is a function of the final-state momenta
$p_i$ ($i=1,2,..,6$) and SM and NP couplings, and is defined with respect
to $dM_2^2 \, dM_5^2 \, d\Omega_1^{**} \, d\Omega_2^{*} \,
d\Omega_4^{**} \, d\Omega_5^{*} \, d\Omega_t$. The cross section is
then integrated over $M_5^2$ and over all angles except for
$\theta_2^*$ and $\theta_\ell^*$. The observables are obtained by (i)
assigning the $p_i$ to specific final-state particles, and (ii)
integrating further over $\theta_2^*$ and $\theta_\ell^*$, or $M_2^2$.

There are three possibilities for the particle assignments: (i) $p_1 =
p_c$, $p_2 = p_b$, $p_3 = p_{\bbar_1}$, (ii) $p_1 = p_c$, $p_2 =
p_{\bbar_1}$, $p_3 = p_b$, (iii) $p_1 = p_b$, $p_2 = p_{\bbar_1}$,
$p_3 = p_c$. Here $p_{\bbar_1}$ refers to the $\bbar$ coming from the
$t$. Also, $p_6 = p_{\ell^-}$. For each case there are two
observables:
\begin{enumerate}

\item {\bf Invariant mass-squared distribution:} 
\begin{eqnarray}
  \frac{d\sigma}{d\zeta_{12}^2} 
  & = & \sigma_{\mbox{\scriptsize SM}}
    \Bigg\{
    F_{12} \frac{6 \,h_{\mbox{\scriptsize SM}}^{12}
    \left(\zeta_{12}^2\right)}
    {\left(1-\zeta_W^2\right)^2
      \left(1+2\zeta_W^2\right)}\nonumber\\
  && ~~~~~~~~~~~~~+ \frac{3 G_Fm_t^2}
     {\sqrt{2}\pi^2 \left(1-\zeta_W^2\right)^2
      \left(1+2\zeta_W^2\right)}
    \sum_{i,\sigma} \hat{A}_i^\sigma h_i^{12}\left(\zeta_{12}^2\right)
\Bigg\},
\label{eq:dsigdM}
\end{eqnarray}

\item {\bf Angular correlation:} 
\begin{eqnarray}
\label{eq:dsigdcosdcos}
  \frac{d\sigma}{d\!\cos\theta_3^* 
    \,d\!\cos\theta_\ell^*}  &=&\frac{\sigma_{\mbox{\scriptsize SM}}}{4}
    \Bigg\{
%    F_{12} 
          \left[1 + \rho_3(\zeta_W^2) \kappa(r)
\cos\theta_3^*\cos\theta_\ell^*\right]\nonumber\\
  &+&  \frac{3 G_F m_t^2}
     {4\sqrt{2}\pi^2 \left(1-\zeta_W^2\right)^2
      \left(1+2\zeta_W^2\right)}\Bigg[
    \Bigg(\sum_{i,\sigma} \hat{A}_i^\sigma\Bigg) \\
  &+& \Bigg(\hat{A}_3^+-\hat{A}_3^-
    -\frac{1}{3}\Big(
      \hat{A}_1^+-\hat{A}_1^-+\hat{A}_2^+-\hat{A}_2^-
    \Big)\Bigg)\kappa(r)
     \cos\theta_3^*\cos\theta_\ell^*\Bigg] \Bigg\}. \nn
%\label{eq:dsigdcosdcos}
\end{eqnarray}

\end{enumerate}
The numerical subscripts and superscripts correspond to the particle
assignments in each of the three cases.  That is, in case (i), the
subscript $12$ corresponds to $bc$ ($b$ is particle 1, $c$ is particle
2), and similarly for cases (ii) ($12 = \bbar c$) and (iii) ($12 =
b\bbar $).  In the summations, $\sigma = +$, $-$ and $i=b$,
$\overline{b}$, $c$.

$\sigma_{SM}$ is given in Eq.~(60) of the Appendix of
Ref.~\cite{companion}, $\zeta_{12}^2 \equiv
\left(p_1+p_2\right)^2/m_t^2$, $\zeta_W \equiv m_W/m_t$, and
$\kappa(r)$ is defined as
\begin{eqnarray}
  \kappa(r) = \frac{\left(-31 r^4 + 37 r^2 -66\right)r 
    -2\left(r^6-17r^4 + 33 r^2-33\right)\tanh^{-1}\left(r\right)}
        {r^2\left[\left(31 r^2 -59\right)r 
          +2\left(r^4-18r^2 + 33\right)\tanh^{-1}\left(r\right)\right]} ~,
\end{eqnarray}
where 
\beq
  r \equiv \sqrt{1 - 4 m_t^2/Q^2} ~,~~ Q \equiv p_t+p_{\tbar} ~.
\eeq
The functions $h_i^{mn}$ ($mn = bc,~\bbar
c,~b\bbar$; $i=b$, $\overline{b}$, $c$) are defined in
Table~\ref{tab:hs}, and
\bea
h_{\mbox{\scriptsize SM}}^{bc}\left(\zeta_{bc}^2\right) 
  & = & \left(1-\zeta_{bc}^2\right)\zeta_{bc}^2\,\,\theta
        (1-\zeta_W^2-\zeta_{bc}^2) ~, \nn\\
h_{\mbox{\scriptsize SM}}^{\bbar c}\left(\zeta_{\bbar c}^2\right) 
  & = & \left(\frac{\zeta_W\gamma_W}{6\pi}\right)
        \frac{(1-\zeta_{\bbar c}^2)^2(1+2\zeta_{\bbar c}^2)}
        {(\zeta_{\bbar c}^2-\zeta_W^2)^2
          +(\zeta_W\gamma_W)^2} ~, \nn\\
  h_{\mbox{\scriptsize SM}}^{b\bbar}
        \left(\zeta_{b\bbar}^2\right) & = &
        \left(1-\zeta_W^2-\zeta_{b\bbar}^2\right)
         \left(\zeta_W^2+\zeta_{b\bbar }^2\right) 
         \theta\!\left(1-\zeta_W^2-\zeta_{b\bbar}^2\right) ~.
\eea

% % % % % % % % % % % % % % % % % % % % % % % % % % % % % % % % % % % 
% % % % % % % % % % % % % % % % % % % % % % % % % % % % % % % % % % % 
% % % % % % % % % % % % % % % % % % % % % % % % % % % % % % % % % % % 
% Table I
\begin{table}[ht]
\bgroup
\def\arraystretch{1.2}
\caption{Definitions of the $h_i^{mn}$ ($mn = bc,~\bbar c,~b\bbar$)
  functions.  The columns correspond to $i=b$, $\overline{b}$, $c$.}
\begin{tabular}{c|ccc}
\hline\hline
& $b$ & $\overline{b}$ & $c$ \\
\hline
 $h_i^{bc}(\zeta^2)$ & ~$\frac{1}{2}(1-\zeta^2)^2(1+2\zeta^2)$~ &
$3(1-\zeta^2)^2\zeta^2$ 
     & ~$\frac{1}{2}(1-\zeta^2)^2(1+2\zeta^2)$~ \\
\hline
 $h_i^{\bbar c}(\zeta^2)$ & $3(1-\zeta^2)^2\zeta^2$ &
~$\frac{1}{2}(1-\zeta^2)^2(1+2\zeta^2)$~
     & ~$\frac{1}{2}(1-\zeta^2)^2(1+2\zeta^2)$~ \\ 
\hline
 $h_i^{b\bbar}(\zeta^2)$ & ~$\frac{1}{2}(1-\zeta^2)^2(1+2\zeta^2)$~ &
$\frac{1}{2}(1-\zeta^2)^2(1+2\zeta^2)$ 
     & ~$3(1-\zeta^2)^2\zeta^2$ ~ \\ 
\hline\hline
\end{tabular}
\label{tab:hs}
\egroup
\end{table}
% % % % % % % % % % % % % % % % % % % % % % % % % % % % % % % % % % % 
% % % % % % % % % % % % % % % % % % % % % % % % % % % % % % % % % % % 
% % % % % % % % % % % % % % % % % % % % % % % % % % % % % % % % % % % 

\noindent 	
In addition,
\bea
\rho_\bbar(\zeta_W^2) = 1 ~,~~
\rho_b(\zeta_W^2) = -\left(\frac{1-2\zeta_W^2}{1+2\zeta_W^2}\right) ~, 
~~~~ \nn\\
\rho_c(\zeta_W^2) = 
           \frac{1-12\zeta_W^2+9\zeta_W^4+2\zeta_W^6-12\zeta_W^4
           \ln(\zeta_W^2) }
           {(1-\zeta_W^2)^2(1+2\zeta_W^2)} ~.
\eea
and
\beq
F_{bc} = F_{b\bbar} = 1 ~,~~ 
F_{\bbar c} = 1 - 4(1 - \zeta_{\bbar c}^2/\zeta_W^2)
\mbox{Re}\left(X^{V}_{LL}\right) ~.
\eeq
Note that we have neglected some mild dependence on
Re$\left(X^V_{LL}\right)$ in the $bc$ and $b\bbar$ distributions.
This dependence is, however, properly taken into account in the
numerical work below.

The NP parameters appear in the observables in the 
$\hat{A}_i^\sigma$'s:
\begin{eqnarray}
  \hat{A}_{\bbar}^+ &=& 4 \left|X^{V}_{LL}\right|^2
  -8 \,\mbox{Re}\left(X^T_{LL}X^{S*}_{LL}\right)+32 
  \left|X^T_{LL}\right|^2 ~,
  \nn\\
  \hat{A}_{\bbar}^- &=&
    4\left|X^{V}_{RR}\right|^2
    -8 \,\mbox{Re}\left(X^T_{RR}X^{S*}_{RR}\right)+32 
    \left|X^T_{RR}\right|^2 ~, \nn\\
  \hat{A}_{b}^+ &=& 
    \left|X^{S}_{LL}\right|^2+\left|X^{S}_{LR}\right|^2
    -16\left|X^{T}_{LL}\right|^2 ~, \nn\\
  \hat{A}_{b}^- &=& 
    \left|X^{S}_{RR}\right|^2+\left|X^{S}_{RL}\right|^2
    -16\left|X^{T}_{RR}\right|^2 ~,\nn\\
  \hat{A}_{c}^+ &=& 
    4\left|X^{V}_{LR}\right|^2
    +8 \,\mbox{Re}\left(X^T_{LL}X^{S*}_{LL}\right)+
    32\left|X^T_{LL}\right|^2 ~, \nn\\
  \hat{A}_{c}^- &=& 
    4\left|X^{V}_{RL}\right|^2
    +8 \,\mbox{Re}\left(X^T_{RR}X^{S*}_{RR}\right)+
    32\left|X^T_{RR}\right|^2 ~.
\label{eq:Ahatdefs}
\end{eqnarray}
As pointed out in the introduction, the six observables have different
functional dependencies on the $\hat{A}_i^\sigma$'s.

%%%%%%%%%%%%%%%%%%%%%%%%%%%%%%%%%%%%%%%%%%%%%%%%%%%%%%%%%%%%%%%%%%%%%%
\section{Extracting New Physics from Observables}
%%%%%%%%%%%%%%%%%%%%%%%%%%%%%%%%%%%%%%%%%%%%%%%%%%%%%%%%%%%%%%%%%%%%%%

In the companion paper \cite{companion}, we showed that the chosen
observables, namely the three invariant mass-squared distributions
($d\sigma/d\zeta^2_{12}$) and the three angular correlations
($d\sigma/d\cos\theta_3^*\cos\theta_{\ell}^*$), are sensitive to
certain combinations of the new-physics parameters. This dependence is
represented in Eqs.~(\ref{eq:dsigdM}) and (\ref{eq:dsigdcosdcos}) as
combinations of the various $\hat{A}_i^\sigma$'s, which in turn can be
defined [Eq.~(\ref{eq:Ahatdefs})] in terms of the NP coefficients 
$X^I_{AB}$ that appear in the effective Lagrangian. 
In addition, the NP operator proportional to $X^V_{LL}$ has the same
Lorentz structure as the corresponding SM operator, so that these two
interfere, leading to an explicit dependence on $X^V_{LL}$ itself.  
As discussed in Ref.~\cite{companion}, the shapes of the observables 
are largely insensitive to effects due to parton distribution 
functions (PDFs). It is only natural then to expect that it should be
possible to extract the various combinations of NP parameters by 
fitting these distributions.  Further, one expects that, by combining
information from all six observables, it should be possible to 
extract the values of the individual $\hat{A}_i^\sigma$'s and
Re($X^V_{LL}$) as well.  In this section we develop the procedure to
carry out this extraction.

\subsection{Algorithm}
\label{sec:algorithm}

Consider first the conventional fitting method, which involves
individual observables. If the underlying theory has $N$ theoretical
unknowns, their values can be determined only if at least $N$
observables are measured. There are theoretical expressions for these
observables in terms of the $N$ unknowns. Using these expressions, the
best-fit values of the unknowns are those for which the measured
values of the observables are best reproduced.

In our case, the observables are distributions and correlations,
specifically $d\sigma/d\zeta^2_{12}$ [Eq.~(\ref{eq:dsigdM})] and
$d\sigma/d\cos\theta_3^*\cos\theta_{\ell}^*$
[Eq.~(\ref{eq:dsigdcosdcos})]. Each distribution/correlation contains
many measurements (at different values of $\zeta^2_{12}$ or
($\cos\theta_3^*$,$\cos\theta_{\ell}^*$)).
Equations~(\ref{eq:dsigdM}) and (\ref{eq:dsigdcosdcos}) show that each
of the $\zeta^2_{12}$ distributions and the angular correlations can
be written as a linear combination of the SM piece and several NP
pieces [six $\hat A_i^{\sigma}$'s, Re($X^V_{LL}$)].  With this in
mind, we use \madg in conjunction with FeynRules~\cite{FeynRules} to
generate eight templates for each distribution/correlation. The
templates are nothing but the said distributions/correlations
generated with the input values of the parameters chosen such that
certain specific contributions to the observables are retained, while
all others are set to zero. The objective is to isolate the
contributions coming from the SM, the individual $\hat A_i^{\sigma}$'s
and Re($X^V_{LL}$).  Table~\ref{tab:templates} gives the parameter
choices made for each template (labelled \tm{i}) and the contributions
that they represent.

Once we have these templates, a $\zeta^2_{12}$ distribution
or an angular correlation arising from a generic choice of NP
parameters can be represented as a linear combination of the
corresponding templates with appropriate coefficients. Extracting
these coefficients allows one to determine the values of the NP
parameters involved.

% % % % % % % % % % % % % % % % % % % % % % % % % % % % % % % % % % % 
% % % % % % % % % % % % % % % % % % % % % % % % % % % % % % % % % % % 
% % % % % % % % % % % % % % % % % % % % % % % % % % % % % % % % % % % 
% Table II
\begin{center}
\begin{table}[ht]
\caption{NP parameter choices for each of the templates \tm{i}.}
\renewcommand{\arraystretch}{1.5}
\renewcommand{\tabcolsep}{0.2cm}
\begin{tabular}{|c|c|c|c|}
\hline 
Template & 
$X^I_{AB}$ & 
$\hat A_i^{\sigma}$ & 
Description \\
\hline \hline
\tm{0} & 
All $X^I_{AB}$ = 0 & 
All $\hat A_i^{\sigma}$ = 0 & 
SM contribution \\
\hline
\tm{1} & 
$X^S_{LL}$, $X^T_{LL}$ $\neq$ 0 & 
$\hat A_{\bbar}^{+}$ $\neq$ 0; all other $\hat A_i^{\sigma}$ = 0 &
Contribution $\propto \hat A_{\bbar}^{+}$ \\
\hline
\tm{2} & 
$X^V_{RR}$ $\neq$ 0 &  
$\hat A_{\bbar}^{-}$
$\neq$ 0; all other $\hat A_i^{\sigma}$ = 0 & 
Contribution $\propto \hat A_{\bbar}^{-}$ \\
\hline
\tm{3} & 
$X^S_{LL}$ $\neq$ 0 &  
$\hat A_{b}^{+}$ $\neq$ 0; all other $\hat A_i^{\sigma}$ = 0 &
Contribution $\propto \hat A_{b}^{+}$ \\
\hline
\tm{4} & 
$X^S_{RR}$ $\neq$ 0 &  
$\hat A_{b}^{-}$ $\neq$ 0; all other $\hat A_i^{\sigma}$ = 0 &
Contribution $\propto \hat A_{b}^{-}$ \\
\hline
\tm{5} & 
$X^V_{LR}$ $\neq$ 0 &  
$\hat A_{c}^{+}$ $\neq$ 0; all other $\hat A_i^{\sigma}$ = 0 &
Contribution $\propto \hat A_{c}^{+}$ \\
\hline
\tm{6} & 
$X^V_{RL}$ $\neq$ 0 & 
$\hat A_{c}^{-}$
$\neq$ 0; all other $\hat A_i^{\sigma}$ = 0 & 
Contribution $\propto \hat A_{c}^{-}$ \\
\hline
\tm{7} & 
$X^V_{LL}$ $\neq$ 0 & 
$\hat A_{\bbar}^{+}$ $\neq$ 0; all other $\hat A_i^{\sigma}$ = 0 &
Contributions $\propto$ Re($X^V_{LL}$) and $\hat A_{\bbar}^{+}$ \\
\hline
\tm{8} & 
\multicolumn{2}{c|}{\tm7 $-$ \tm1} & 
Contribution $\propto$ Re($X^V_{LL}$) \\
\hline
\end{tabular}
\label{tab:templates}
\end{table}
\end{center}
% % % % % % % % % % % % % % % % % % % % % % % % % % % % % % % % % % % 
% % % % % % % % % % % % % % % % % % % % % % % % % % % % % % % % % % % 
% % % % % % % % % % % % % % % % % % % % % % % % % % % % % % % % % % % 

\vspace*{-40pt}
\subsection{Testing the algorithm}

We use ``pseudo-data'' generated in Monte Carlo
simulations to test our fitting procedure. Once again, we use 
\madg to generate these
samples. In Ref.~\cite{companion} we presented plots of the normalized
distributions and correlations. Here we use the unnormalized
distributions for the fitting. Since both the templates and the
``data-sets'' are obtained with \madg with the same choices of PDFs,
scale, etc., the overall normalization is automatically accounted for.

Our procedure is as follows. We generate pseudo-data using MadGraph 5,
in conjunction with FeynRules, for certain chosen values of the NP
parameters (i.e., the $X^I_{AB}$). This gives us three $\zeta^2_{12}$
distributions and three angular correlations. We divide each of these
into 25 bins (using a $5\times 5$ array for the angular correlations).
We then perform a single $\chi^2$ minimization involving all six
histograms in order to determine the coefficients for the templates
that result in the best fit for all six observables
simultaneously. For this purpose, we use standard, publicly-available
routines \cite{NRecipes}. Finally, we examine to what extent the
values of the NP parameters extracted from the fit agree with their
input values.

We consider four different test cases of pseudo-data, which we label
\ex{1}, \ex{2}, \ex{3} and \ex{4}. The values of the input NP
parameters for these test cases are listed in
Table~\ref{tab:testcases}, along with the size of the cross section
relative to the SM prediction. The data sets \ex{i} have
been generated for the process $\ggprocessenu$, taking a benchmark
luminosity that corresponds to $10^5$ SM events. The uncertainties
incorporated in the fitting procedure are statistical only, and are
estimated by considering the number of events in each bin in the
histograms to be Poisson-distributed. The
templates \tm{i} have been generated for the same process
but with ${\cal O}$(10$^6$) events, so that uncertainties from these 
can be neglected in the fit.

% % % % % % % % % % % % % % % % % % % % % % % % % % % % % % % % % % % 
% % % % % % % % % % % % % % % % % % % % % % % % % % % % % % % % % % % 
% % % % % % % % % % % % % % % % % % % % % % % % % % % % % % % % % % % 
% Table III
\begin{center}
\begin{table}[ht]
\caption{Input values of the NP parameters for the four test cases
  \ex{i}. The last column illustrates how the total cross section
  $\sigma$ is affected in each of the test cases.}
\renewcommand{\arraystretch}{1.5} \renewcommand{\tabcolsep}{0.2cm}
\begin{tabular}{|c|l|l|c|}
\hline 
Test Case & 
\multicolumn{1}{c|}{$X^I_{AB}$} & 
\multicolumn{1}{c|}{$\hat A_i^{\sigma}$} &
$\sigma / \sigma_{SM}$ \\
\hline \hline
%#####################################################################
\texttt{EX-1} & 
$X^T_{LL}$ = 1 ; $X^T_{RR}$ = 1 & 
$\hat A_{\bbar}^{+}$ = 32 ; 
$\hat A_{b}^{+}$ = $-$16 ;
$\hat A_{c}^{+}$ = 32 ; &
3.1 \\
 &
 &
$\hat A_{\bbar}^{-}$ = 32 ; 
$\hat A_{b}^{-}$ = $-$16 ; 
$\hat A_{c}^{-}$ = 32 &
 \\
\hline
%#####################################################################
\texttt{EX-2} & 
$X^S_{LR}$ = 5 & 
$\hat A_{b}^{+}$ = 25 ; all other $\hat A_i^{\sigma}$'s = 0 &
1.6 \\
\hline
%#####################################################################
\texttt{EX-3} & 
$X^S_{LR}$ = 3 ; $X^S_{RL}$ = 4 & 
$\hat A_{b}^{+}$ = 9 ; 
$\hat A_{b}^{-}$ = 16 ; 
all other $\hat A_i^{\sigma}$'s = 0 &
1.6 \\
\hline
%#####################################################################
\texttt{EX-4} & 
$X^V_{LL}$ = 3 ; $X^S_{LL}$ = 5 & 
$\hat A_{\bbar}^{+}$ = 36 ; 
$\hat A_{b}^{+}$ = 25 ;  
all other $\hat A_i^{\sigma}$'s = 0 &
2.4 \\
\hline
\end{tabular}
\label{tab:testcases}
\end{table}
\end{center}
% % % % % % % % % % % % % % % % % % % % % % % % % % % % % % % % % % % 
% % % % % % % % % % % % % % % % % % % % % % % % % % % % % % % % % % % 
% % % % % % % % % % % % % % % % % % % % % % % % % % % % % % % % % % % 

\subsubsection{\textbf{Fit 1}}

As detailed above, the templates are generated by assuming there is
only a single contribution at a time to the distributions/correlations
for the process $\ggprocessenu$. The observables are represented by
the analytical expressions in Eqs.~(\ref{eq:dsigdM}) and
(\ref{eq:dsigdcosdcos}). However, these expressions have been derived
\cite{companion} under the (unrealistic) assumption that the two
$\bbar$'s in the final state are distinguishable. In Fit 1, as a first
test of the algorithm, we retain this assumption.

Table~\ref{tab:fit1} shows the values of the $\hat A_i^{\sigma}$'s and
Re($X^V_{LL}$) extracted from the fit for the four different test
cases of pseudo-data in Table~\ref{tab:testcases}. A comparison of the
two tables shows that most of the values of the parameters extracted
from the fit agree with their input values within $\pm 1\sigma$.  This
demonstrates that the fundamental idea of the algorithm, namely
fitting using the templates, is sound.

The worst-fitted parameter is Re($X^V_{LL}$) in the case where $\hat
A_{\bar b}^+$ is nonzero but Re($X^V_{LL}) = 0$. This poor fit is an
artifact of the somewhat simple-minded fitting procedure that we
adopt: $\hat A_{\bar b}^+$ and Re($X^V_{LL}$) are treated as
independent parameters in the fit, despite the fact that they are
correlated [see Eq.~(\ref{eq:Ahatdefs})]. Note that the contribution
proportional exclusively to Re($X^V_{LL}$) appears primarily in
$d\sigma/d\zeta^2_{\bbar c}$.\footnote{Based on our theoretical
  analysis, we expect the $\zeta_{\bbar c}^2$ distribution to have the
  most sensitivity to Re($X^V_{LL}$). This expectation is confirmed by
  an examination of the templates.  Having said this, the dependence 
  on Re($X^V_{LL}$) is not completely negligible for the other
  distributions and correlations, and in our numerical work
  we include the corresponding template (\tm{8}) in the fits for all
  distributions and correlations.} Even so, the fit performs
rather well when Re($X^V_{LL}$) is, in fact, nonzero. On the other
hand, not considering Re($X^V_{LL}$) as a fit parameter leads to an
overall worsening of the fits.  For this reason we retain it in our
fitting algorithm, while taking care to avoid drawing any strong
conclusions from the extracted value of this parameter.

% % % % % % % % % % % % % % % % % % % % % % % % % % % % % % % % % % % 
% % % % % % % % % % % % % % % % % % % % % % % % % % % % % % % % % % % 
% % % % % % % % % % % % % % % % % % % % % % % % % % % % % % % % % % % 
% Table IV
% Fit for all observables
% initial state - gg
% crossed-diagrams - not included
{
\renewcommand{\arraystretch}{1.5}
\renewcommand{\tabcolsep}{0.2cm}
\begin{longtable}{|c|l@{}l|c|}
\caption{Values of the NP parameters extracted from the four test 
cases \ex{i} using Fit 1.} \\
\hline
Test Case & 
\multicolumn{2}{c|}{Fit Results} &
$\chi^2/{\rm d.o.f.}$ \\
\hline \hline 
\endfirsthead

\multicolumn{4}{c}
{{\tablename\ \thetable{} -- continued }} \\
\hline
Test Case &
\multicolumn{2}{c|}{Fit Results} &
$\chi^2/{\rm d.o.f.}$ \\ 
\hline
\endhead

\hline 
\endfoot

\hline
\endlastfoot

%#####################################################################
%Fit for EX-1
%#####################################################################
%---------------------------------------------------------------------
\texttt{EX-1} & 
SM coeff. = 1.005 $\pm$ 0.003 &
 &
1.30 \\
%---------------------------------------------------------------------
 & 
$\hat A_{\bbar}^{+}$ = 33 $\pm$ 2 &  
$\hat A_{\bbar}^{-}$ = 30 $\pm$ 2 &
 \\
%---------------------------------------------------------------------
 &
$\hat A_{b}^{+}$ = $-$16 $\pm$ 2 &
$\hat A_{b}^{-}$ = $-$15 $\pm$ 2 &
 \\
%---------------------------------------------------------------------
 &
$\hat A_{c}^{+}$ = 33 $\pm$ 2 &
$\hat A_{c}^{-}$ = 31 $\pm$ 2 &
 \\
%---------------------------------------------------------------------
 &
\multicolumn{2}{l|}{Re($X^V_{LL}$) = 0.40 $\pm$ 0.02} &
 \\
%---------------------------------------------------------------------
\hline\hline
%#####################################################################
%#####################################################################
%
%
%#####################################################################
%Fit for EX-2
%#####################################################################
%---------------------------------------------------------------------
\texttt{EX-2} & 
SM coeff. = 1.000 $\pm$ 0.002 &
 &
1.21 \\ 
%---------------------------------------------------------------------
 & 
$\hat A_{\bbar}^{+}$ = 0 $\pm$ 1 &  
$\hat A_{\bbar}^{-}$ = 0 $\pm$ 1 &
 \\
%---------------------------------------------------------------------
 &
$\hat A_{b}^{+}$ = 24 $\pm$ 1 &
$\hat A_{b}^{-}$ = 1 $\pm$ 1 &
 \\
%---------------------------------------------------------------------
 &
$\hat A_{c}^{+}$ = 1 $\pm$ 1 &
$\hat A_{c}^{-}$ = 0 $\pm$ 1 &
 \\
 &
\multicolumn{2}{l|}{Re($X^V_{LL}$) = 0.01 $\pm$ 0.02} &
 \\
\hline\hline
%#####################################################################
%#####################################################################
%
%
%#####################################################################
%Fit for EX-3
%#####################################################################
%---------------------------------------------------------------------
\texttt{EX-3} & 
SM coeff. = 0.994 $\pm$ 0.002 &
 &
1.22 \\ 
%---------------------------------------------------------------------
 & 
$\hat A_{\bbar}^{+}$ = 1 $\pm$ 1 &  
$\hat A_{\bbar}^{-}$ = $-$1 $\pm$ 1 &
 \\
%---------------------------------------------------------------------
 &
$\hat A_{b}^{+}$ = 10 $\pm$ 1 &
$\hat A_{b}^{-}$ = 15 $\pm$ 1 &
 \\
%---------------------------------------------------------------------
 &
$\hat A_{c}^{+}$ = 0 $\pm$ 1 &
$\hat A_{c}^{-}$ = 0 $\pm$ 1 &
 \\
%---------------------------------------------------------------------
 &
\multicolumn{2}{l|}{Re($X^V_{LL}$) = 0.02 $\pm$ 0.02} &
 \\
%---------------------------------------------------------------------
\hline\hline
%#####################################################################
%#####################################################################
%
%
%#####################################################################
%Fit for EX-4
%#####################################################################
%---------------------------------------------------------------------
\texttt{EX-4} & 
SM coeff. = 1.003 $\pm$ 0.003 &
 &
1.43 \\
%---------------------------------------------------------------------
 & 
$\hat A_{\bbar}^{+}$ = 36 $\pm$ 1 &  
$\hat A_{\bbar}^{-}$ = 0 $\pm$ 1 &
 \\
%---------------------------------------------------------------------
 &
$\hat A_{b}^{+}$ = 25 $\pm$ 1 &
$\hat A_{b}^{-}$ = $-$1 $\pm$ 1 &
 \\
%---------------------------------------------------------------------
 &
$\hat A_{c}^{+}$ = 1 $\pm$ 1 &
$\hat A_{c}^{-}$ = $-$1 $\pm$ 1 &
 \\
%---------------------------------------------------------------------
 &
\multicolumn{2}{l|}{Re($X^V_{LL}$) = 3.03 $\pm$ 0.01} &
%---------------------------------------------------------------------
%#####################################################################
%#####################################################################

\label{tab:fit1}
\end{longtable}
}
% % % % % % % % % % % % % % % % % % % % % % % % % % % % % % % % % % % 
% % % % % % % % % % % % % % % % % % % % % % % % % % % % % % % % % % % 
% % % % % % % % % % % % % % % % % % % % % % % % % % % % % % % % % % % 

\subsubsection{\textbf{Fit 2}}

In Fit 2 we drop the assumption that the two final-state $\bbar$'s are
distinguishable. The Monte Carlo pseudo-data (as well as the
templates) for the process $\ggprocessenu$ now includes the amplitudes
in which the momenta of the two $\bbar$'s in the final state are
exchanged. However, in order to construct the above observables, we
necessarily need to identify the $\bar b$ emerging from the top decay.
Hence we must restrict our analysis to regions of phase space where
the two $\bar b$'s can effectively be considered to be
distinguishable.  To do this, we construct the two quantities $m_1^2 =
(p_b + p_c + p_{\bbar_1})^2$ and $m_2^2 = (p_b + p_c + p_{\bbar_2})^2
$. If both $m_1$ and $m_2$ lie within the range $m_t \pm 15\Gamma_t$,
the event is discarded.  Otherwise, it is assumed that the $\bbar_i$
that yields the smaller value of $|m_i - m_t|$ comes from the
$t$-quark decay. This leads to a loss of about 20\% of the
events. This cut also distorts the angular correlation such that it no
longer conforms to the familiar 
($a_1 + a_2\cos\theta_3^*\cos\theta_\ell^*$) form, even
for the SM. Despite the distortion, the fit can be performed using the
same algorithm as long as the the same method of event selection is
applied to the pseudo-data as well as the templates.

The results of the fit are presented in Table~\ref{tab:fit2} for the
four test cases. The agreement between the values of the fitted
parameters and their input values is almost as good as in the
idealized case (Fit 1): apart from Re($X^V_{LL}$) in \ex{1}, all
values agree within $\pm 1.5\sigma$. We find that, with the event
selection 
discussed above, despite the resulting loss of statistics, one obtains 
a slight improvement in the goodness-of-fit, as can be seen from the
smaller values of $\chi^2/{\rm d.o.f.}$ 
All of this demonstrates that our algorithm
continues to hold, even when one imposes a `cut' to distinguish the
two $\bbar$'s in the final state.

% % % % % % % % % % % % % % % % % % % % % % % % % % % % % % % % % % % 
% % % % % % % % % % % % % % % % % % % % % % % % % % % % % % % % % % % 
% % % % % % % % % % % % % % % % % % % % % % % % % % % % % % % % % % % 
% Table V
% Fit for all observables
% initial state - gg
% crossed-diagrams - included
{
\renewcommand{\arraystretch}{1.5}
\renewcommand{\tabcolsep}{0.2cm}
\begin{longtable}{|c|l@{ ; }l|c|}
\caption{Values of the NP parameters extracted from the four test 
cases \ex{i} using Fit 2.} \\
\hline
Test Case & 
\multicolumn{2}{c|}{Fit Results} &
$\chi^2/{\rm d.o.f.}$ \\
\hline \hline 
\endfirsthead

\multicolumn{4}{c}
{{\tablename\ \thetable{} -- continued }}\\
\hline
Test Case &
\multicolumn{2}{c|}{Fit Results} &
$\chi^2/{\rm d.o.f.}$ \\ 
\hline
\endhead

\hline 
\endfoot

\hline
\endlastfoot

%#####################################################################
%Fit for EX-1
%#####################################################################
%---------------------------------------------------------------------
\texttt{EX-1} & 
SM coeff. = 1.005 $\pm$ 0.004 &
 &
1.09 \\
%---------------------------------------------------------------------
 & 
$\hat A_{\bbar}^{+}$ = 29 $\pm$ 2 &  
$\hat A_{\bbar}^{-}$ = 33 $\pm$ 2 &
 \\
%---------------------------------------------------------------------
 &
$\hat A_{b}^{+}$ = $-$15 $\pm$ 2 &
$\hat A_{b}^{-}$ = $-$16 $\pm$ 2 &
 \\
%---------------------------------------------------------------------
 &
$\hat A_{c}^{+}$ = 33 $\pm$ 2 &
$\hat A_{c}^{-}$ = 31 $\pm$ 2 &
 \\
%---------------------------------------------------------------------
 &
\multicolumn{2}{l|}{Re($X^V_{LL}$) = 0.40 $\pm$ 0.03} &
 \\
%---------------------------------------------------------------------
\hline\hline
%#####################################################################
%#####################################################################
%
%
%#####################################################################
%Fit for EX-2
%#####################################################################
%---------------------------------------------------------------------
\texttt{EX-2} & 
SM coeff. = 0.999 $\pm$ 0.003 &
 &
1.05 \\
%---------------------------------------------------------------------
 & 
$\hat A_{\bbar}^{+}$ = 1 $\pm$ 1 &  
$\hat A_{\bbar}^{-}$ = $-$1 $\pm$ 1 &
 \\
%---------------------------------------------------------------------
 &
$\hat A_{b}^{+}$ = 26 $\pm$ 1 &
$\hat A_{b}^{-}$ = 0 $\pm$ 1 &
 \\
%---------------------------------------------------------------------
 &
$\hat A_{c}^{+}$ = 1 $\pm$ 2 &
$\hat A_{c}^{-}$ = $-$2 $\pm$ 2 &
 \\
%---------------------------------------------------------------------
 &
\multicolumn{2}{l|}{Re($X^V_{LL}$) = $-$0.01 $\pm$ 0.02} &
 \\
%---------------------------------------------------------------------
\hline\hline
%#####################################################################
%#####################################################################
%
%
%#####################################################################
%Fit for EX-3
%#####################################################################
%---------------------------------------------------------------------
\texttt{EX-3} & 
SM coeff. =  1.005 $\pm$ 0.003 &
 &
0.89 \\
%---------------------------------------------------------------------
 & 
$\hat A_{\bbar}^{+}$ = $-$1 $\pm$ 1 &  
$\hat A_{\bbar}^{-}$ = 1 $\pm$ 1 &
 \\
%---------------------------------------------------------------------
 &
$\hat A_{b}^{+}$ = 10 $\pm$ 1 &
$\hat A_{b}^{-}$ = 15 $\pm$ 1 &
 \\
%---------------------------------------------------------------------
 &
$\hat A_{c}^{+}$ = 0 $\pm$ 2 &
$\hat A_{c}^{-}$ = 0 $\pm$ 2 &
 \\
%---------------------------------------------------------------------
 &
\multicolumn{2}{l|}{Re($X^V_{LL}$) = $-$0.01 $\pm$ 0.02} &
 \\
%---------------------------------------------------------------------
\hline\hline
%#####################################################################
%#####################################################################
%
%
%#####################################################################
%Fit for EX-4
%#####################################################################
%---------------------------------------------------------------------
\texttt{EX-4} & 
SM coeff. = 0.997 $\pm$ 0.003 &
 &
1.29 \\
%---------------------------------------------------------------------
 & 
$\hat A_{\bbar}^{+}$ = 38 $\pm$ 2 &  
$\hat A_{\bbar}^{-}$ = $-$1 $\pm$ 2 &
 \\
%---------------------------------------------------------------------
 &
$\hat A_{b}^{+}$ = 23 $\pm$ 2 &
$\hat A_{b}^{-}$ = 2 $\pm$ 2 &
 \\
%---------------------------------------------------------------------
 &
$\hat A_{c}^{+}$ = 1 $\pm$ 2 &
$\hat A_{c}^{-}$ = $-$1 $\pm$ 2 &
 \\
%---------------------------------------------------------------------
 &
\multicolumn{2}{l|}{Re($X^V_{LL}$) = 2.98 $\pm$ 0.02} &
%---------------------------------------------------------------------
%#####################################################################
%#####################################################################

\label{tab:fit2}
\end{longtable}
}
% % % % % % % % % % % % % % % % % % % % % % % % % % % % % % % % % % % 
% % % % % % % % % % % % % % % % % % % % % % % % % % % % % % % % % % % 
% % % % % % % % % % % % % % % % % % % % % % % % % % % % % % % % % % % 

\subsubsection{\textbf{Fit 3}}

Finally, at the LHC, there is a small ($\approx 10$-15\%) contribution
to $t \tbar $ production from $q \bar q$ annihilation. In Fit 3 we
consider the impact of this additional contribution.

It must be said that we do not expect a significant effect. Since the
NP couplings play a role only in top decay, the structure of
Eqs.~(\ref{eq:dsigdM}) and (\ref{eq:dsigdcosdcos}) remains largely
unchanged. The change in Eq.~(\ref{eq:dsigdM}) is the analytical form
of the factor $\sigma_{SM}$; in Eq.~(\ref{eq:dsigdcosdcos}), the
changes appear in the expressions for $\sigma_{SM}$ and
$\kappa(r)$. The decomposition of the NP contribution in terms of a
linear combination of $\hat A_i^{\sigma}$'s and Re($X^V_{LL}$)
therefore remains valid for the purposes of the fit. Moreover, in the
$\sqrt{\hat s}$ range that is sampled\footnote{At a 14 TeV pp collider,
with ${\cal O}(10^5)$ events, this range is approximately 350 GeV to 
1200 GeV.}, $t \tbar$ production is overwhelmingly dominated by $gg$
fusion, simply because the gluon density is large at low values of
momentum fractions (the well-known Bjorken $x_1$ and $x_2$). This means
that the corrections due to $q \bar q \to t \tbar$ are small in
magnitude over the entire region of phase space that can be probed.

The results of Fit 3 are presented in Table~\ref{tab:fit3}. As
expected, the fitting procedure described above proves just as
effective for the full process $p p \to t \tbar \to (b \bbar c) (\bbar
e^- \bar \nu_e)$.

% % % % % % % % % % % % % % % % % % % % % % % % % % % % % % % % % % % 
% % % % % % % % % % % % % % % % % % % % % % % % % % % % % % % % % % % 
% % % % % % % % % % % % % % % % % % % % % % % % % % % % % % % % % % % 
% Table VI
% Fit for all observables
% initial state - pp
% crossed-diagrams - included
{
\renewcommand{\arraystretch}{1.5}
\renewcommand{\tabcolsep}{0.2cm}
\begin{longtable}{|c|l@{ ; }l|c|}
\caption{Values of the NP parameters extracted from the four test 
cases \ex{i} using Fit 3.} \\
\hline
Test Case & 
\multicolumn{2}{c|}{Fit Results} &
$\chi^2/{\rm d.o.f.}$ \\
\hline \hline 
\endfirsthead

\multicolumn{4}{c}
{{\tablename\ \thetable{} -- continued }} \\
\hline
Test Case &
\multicolumn{2}{c|}{Fit Results} &
$\chi^2/{\rm d.o.f.}$ \\ 
\hline
\endhead

\hline 
\endfoot

\hline
\endlastfoot

%#####################################################################
%Fit for EX-1
%#####################################################################
%---------------------------------------------------------------------
\texttt{EX-1} & 
SM coeff. = 1.002 $\pm$ 0.004 &
 &
1.01 \\
%---------------------------------------------------------------------
 & 
$\hat A_{\bbar}^{+}$ =  32 $\pm$ 3 &  
$\hat A_{\bbar}^{-}$ =  31 $\pm$ 3 &
 \\
%---------------------------------------------------------------------
 &
$\hat A_{b}^{+}$ = $-$16 $\pm$ 2 &
$\hat A_{b}^{-}$ = $-$15 $\pm$ 2 &
 \\
%---------------------------------------------------------------------
 &
$\hat A_{c}^{+}$ =  32 $\pm$ 3 &
$\hat A_{c}^{-}$ =  33 $\pm$ 3 &
 \\
%---------------------------------------------------------------------
 &
\multicolumn{2}{l|}{Re($X^V_{LL}$) = 0.42 $\pm$ 0.03} &
 \\
%---------------------------------------------------------------------
\hline\hline
%#####################################################################
%#####################################################################
%
%
%#####################################################################
%Fit for EX-2
%#####################################################################
%---------------------------------------------------------------------
\texttt{EX-2} & 
SM coeff. = 0.998 $\pm$ 0.003 &
 &
1.00 \\ 
%---------------------------------------------------------------------
 & 
$\hat A_{\bbar}^{+}$ = $-$1 $\pm$ 2 &  
$\hat A_{\bbar}^{-}$ = 0 $\pm$ 2 &
 \\
%---------------------------------------------------------------------
 &
$\hat A_{b}^{+}$ = 24 $\pm$ 2 &
$\hat A_{b}^{-}$ = 1 $\pm$ 2 &
 \\
%---------------------------------------------------------------------
 &
$\hat A_{c}^{+}$ = 1 $\pm$ 2 &
$\hat A_{c}^{-}$ = $-$1 $\pm$ 2 &
 \\
%---------------------------------------------------------------------
 &
\multicolumn{2}{l|}{Re($X^V_{LL}$) = $-$0.01 $\pm$ 0.02} &
 \\
%---------------------------------------------------------------------
\hline\hline
%#####################################################################
%#####################################################################
%
%
%#####################################################################
%Fit for EX-3
%#####################################################################
%---------------------------------------------------------------------
\texttt{EX-3} & 
SM coeff. = 1.001 $\pm$ 0.003 &
 &
1.08 \\
%---------------------------------------------------------------------
 & 
$\hat A_{\bbar}^{+}$ = 1 $\pm$ 2 &  
$\hat A_{\bbar}^{-}$ = $-$1 $\pm$ 2 &
 \\
%---------------------------------------------------------------------
 &
$\hat A_{b}^{+}$ = 9 $\pm$ 2 &
$\hat A_{b}^{-}$ = 16 $\pm$ 2 &
 \\
%---------------------------------------------------------------------
 &
$\hat A_{c}^{+}$ = 1 $\pm$ 2 &
$\hat A_{c}^{-}$ = $-$1 $\pm$ 2 &
 \\
%---------------------------------------------------------------------
 &
\multicolumn{2}{l|}{Re($X^V_{LL}$) = $-$0.01 $\pm$ 0.02} &
 \\
%---------------------------------------------------------------------
\hline\hline
%#####################################################################
%#####################################################################
%
%
%#####################################################################
%Fit for EX-4
%#####################################################################
%---------------------------------------------------------------------
\texttt{EX-4} & 
SM coeff. = 0.999 $\pm$ 0.003 &
 &
1.01 \\
%---------------------------------------------------------------------
 & 
$\hat A_{\bbar}^{+}$ = 38 $\pm$ 2 &  
$\hat A_{\bbar}^{-}$ = $-$2 $\pm$ 2 &
 \\
%---------------------------------------------------------------------  
 &
$\hat A_{b}^{+}$ = 24 $\pm$ 2 &
$\hat A_{b}^{-}$ = 1 $\pm$ 2 &
 \\
%---------------------------------------------------------------------
 &
$\hat A_{c}^{+}$ = 1 $\pm$ 3 &
$\hat A_{c}^{-}$ = $-$1 $\pm$ 2 &
 \\
%---------------------------------------------------------------------
 &
\multicolumn{2}{l|}{Re($X^V_{LL}$) = 2.97 $\pm$ 0.02} &
%---------------------------------------------------------------------
%#####################################################################
%#####################################################################

\label{tab:fit3}
\end{longtable}
}
% % % % % % % % % % % % % % % % % % % % % % % % % % % % % % % % % % % 
% % % % % % % % % % % % % % % % % % % % % % % % % % % % % % % % % % % 
% % % % % % % % % % % % % % % % % % % % % % % % % % % % % % % % % % % 

Although the essential structure of our statistical analysis is based
on the analytical expressions obtained in Ref.~\cite{companion}, where
several simplifying assumptions were made, through the above series of
fits we have obtained a reliable algorithm that includes a procedure
to distinguish the two final-state $\bbar$'s, and works well even in
the presence of the contribution from $q \bar q \to t \tbar$. We now
use this algorithm to examine the prospects for obtaining information
about NP in the decay $\tbbc$ at the LHC, in both the short and long
terms. This is discussed in the next section.

%%%%%%%%%%%%%%%%%%%%%%%%%%%%%%%%%%%%%%%%%%%%%%%%%%%%%%%%%%%%%%%%%%%%%%
\section{Detecting New Physics in Top Decay}
%%%%%%%%%%%%%%%%%%%%%%%%%%%%%%%%%%%%%%%%%%%%%%%%%%%%%%%%%%%%%%%%%%%%%%

Above, we have established a method for the extraction of NP
parameters involved in the decay $\tbbc$. However, should a sizeable
NP contribution exist, it is likely that it would first be detected
simply by measuring the total cross section in this channel. It is
only afterwards that the $\zeta^2_{12}$ distributions and the
angular correlations discussed in the preceding sections would be used
to indicate the presence of NP. While this is true, it should also be
pointed out that the overall normalization of the cross section
suffers from inherent theoretical uncertainties such as the choice of
PDFs, the renormalization and factorization scales, etc. On the other
hand, compared to the total cross section, the
distributions/correlations have additional discriminating power since
their shapes also get modified under the influence of NP.

In the following subsections, using the distributions/correlations, we
perform simulations to examine the prospects for detecting NP, for
measuring certain combinations of NP parameters, and for partially
identifying the NP. The simulations are done using a total number of
events consistent with either short-term or long-term measurements at
the LHC.

\subsection{Short term}

The $t \tbar$ cross section at the LHC at a centre-of-mass energy of
14 TeV is $\sim 900$ pb \cite{topNLO}. Considering the SM branching
fractions for $t \to b \bbar c$ and $\bar t \to \bbar \ell^- \bar
\nu_{\ell}$, the effective cross section in this channel is $\sim 0.1$
pb. For the short-term simulations, we consider an integrated
luminosity which, after factoring in the $b$-tagging
efficiency\footnote{This is assumed to be 70\% for each of the three
  $b$ or $\bbar$'s in the final state}, will lead to $10^4$ events of
the type $p p \to t \tbar \to (b \bbar c) (\bbar \ell^- \bar
\nu_{\ell})$ from the SM alone\footnote{Here $\ell = e,\mu$. In the
  CP-conserving scenario that we consider, there would be an equal
  number of events in which $\bar t \to \bbar b \bar c$ and $t \to b
  \ell^+ \nu_{\ell}$. We assume that the events in which the $\tbar$
  decays leptonically can be identified by tagging the charge of the
  lepton and consider only those events in our analysis.}. This is
expected to be delivered by 2020-2021 \cite{LHCprojection}.

\begin{figure}[!htbp]
\centering
\subfigure[]
{
\includegraphics[scale=0.8]{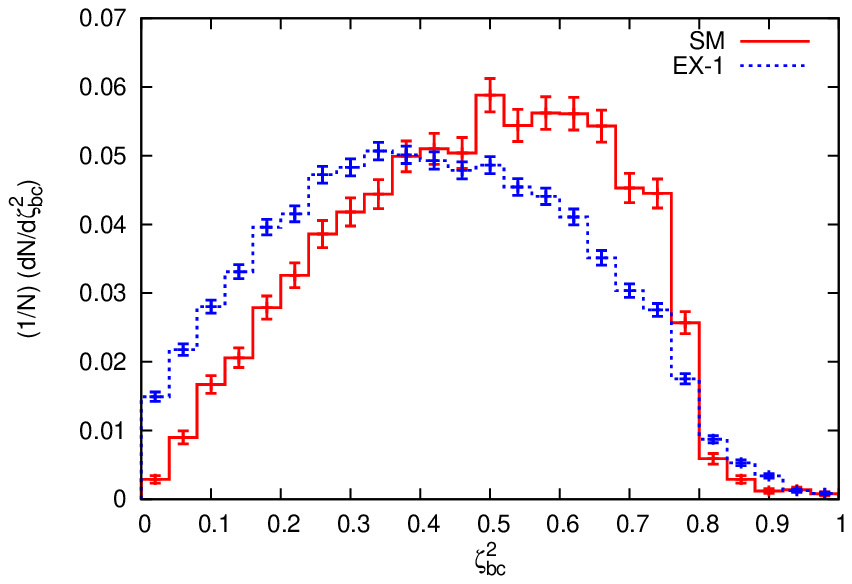}
\label{fig:zetasq_bc}
}
\subfigure[]
{
\includegraphics[scale=0.8]{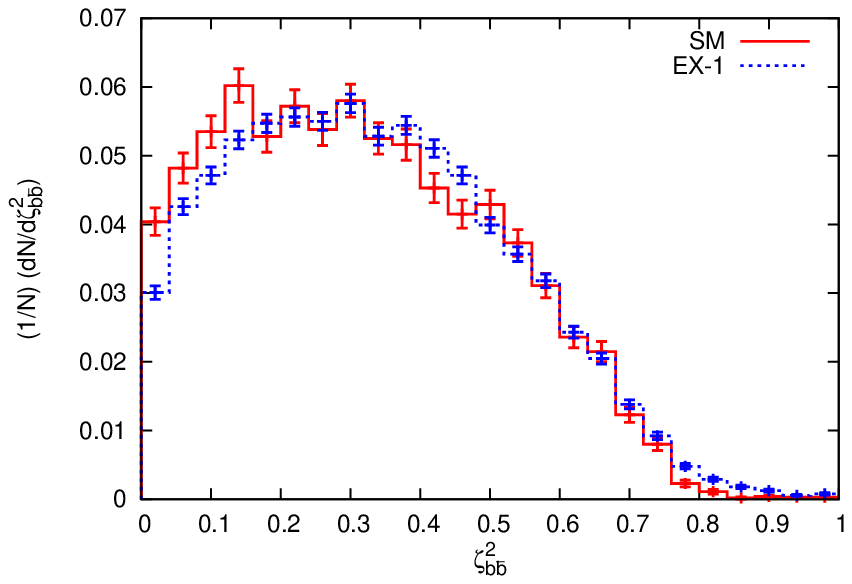}
\label{fig:zetasq_bbbar}
}
\subfigure[]
{
\includegraphics[scale=0.8]{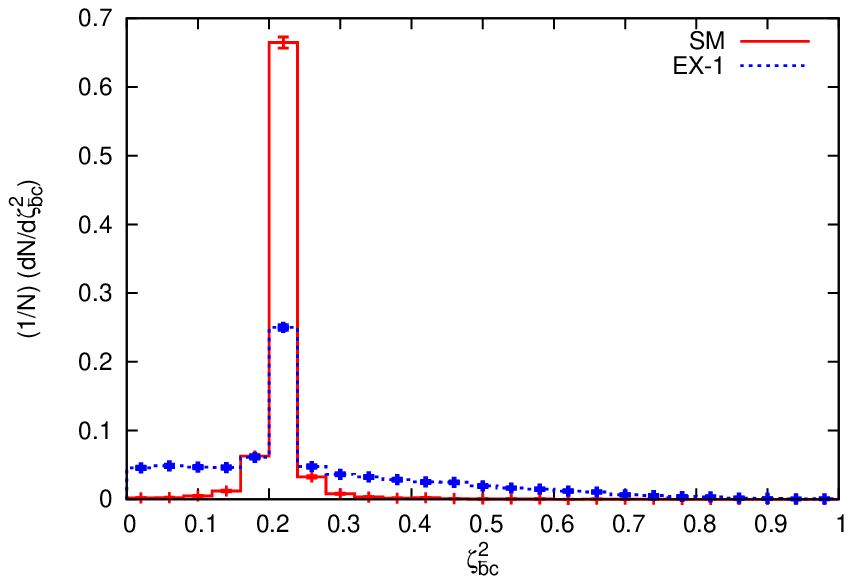}
\label{fig:zetasq_bbarc}
}
\subfigure[]
{
\includegraphics[scale=0.8]{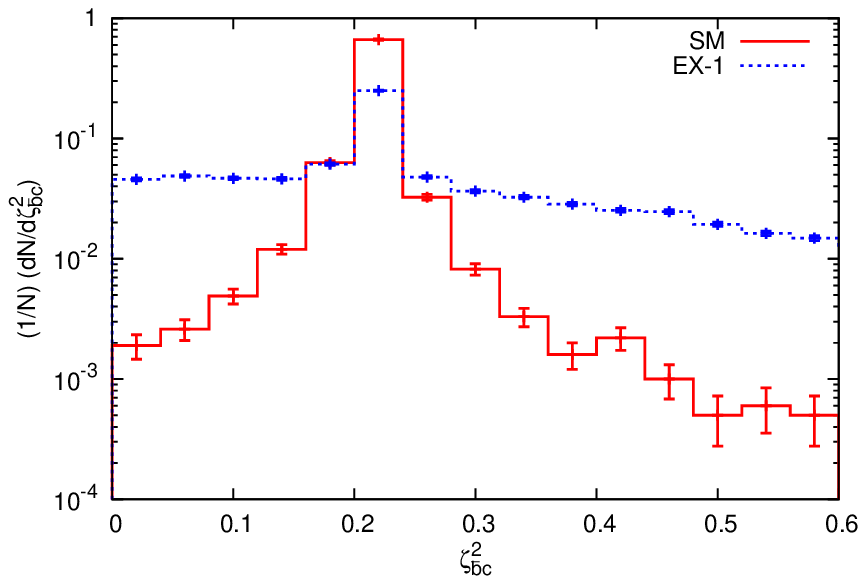}
\label{fig:zetasq_bbarc_semilog}
}
\caption{Normalized $d\sigma/d\zeta^2_{12}$ distributions: (a)
  $d\sigma/d\zeta^2_{bc}$, (b) $d\sigma/d\zeta^2_{b \bar b}$, (c)
  $d\sigma/d\zeta^2_{\bbar c}$, (d) $d\sigma/d\zeta^2_{\bbar c}$ on
  a semi-log scale.}
\label{fig:zetasq}
\end{figure}

In the preceding sections, we noted that, in the presence of NP, the
shapes of the distributions/correlations can be modified. This
suggests that NP can be detected by examining a particular
distribution/correlation and seeing a clear difference between the
measured shape and its SM prediction. This is explored in
Fig.~\ref{fig:zetasq}. Here all three $\zeta^2_{12}$ distributions are
shown for the NP scenario \ex{1}. Clearly, in the cases of
$d\sigma/d\zeta^2_{bc}$ and $d\sigma/d\zeta^2_{\bar b c}$, the
measurement of the distributions alone would indicate the presence of
NP.  On the other hand, it would be difficult to draw conclusions from
the shape of the corresponding $d\sigma/d\zeta^2_{b \bar b}$
distribution.\footnote{If we were not normalizing
  the distribution to the total number of events, the difference between
the SM and \ex{1} cases would be much more apparent.}

However, even in the case of $d\sigma/d\zeta^2_{b \bar b}$,
information about the NP can be obtained. To see this, we use the
fitting procedure developed in the previous section and examine what
kind of information can be extracted by fitting this distribution
alone. From Eq.~(\ref{eq:dsigdM}) one sees that the $\zeta^2_{b \bar
  b}$ distribution depends on three distinct kinematic structures:
$h_{SM}^{b \bbar}$, $h_c^{b \bbar}$ and $h_b^{b \bbar}$ (=
$h_{\bbar}^{b \bbar}$). These kinematic structures will in principle
be modified by cuts, such as we place on the final state $\bbar$'s.
In addition, as described above, there is a mild (but potentially
important) dependence on 
Re$\left(X^V_{LL}\right)$ in the $\zeta^2_{b \bar  b}$ distribution.  
We did not include this dependence in the analytical expressions 
above, but we do retain it here in our numerical work. Thus, a fit
using only the $\zeta^2_{b \bar b}$ distribution would be sensitive to
the relative weights of the SM contribution, Re($X^V_{LL})$, ($\hat
A_c^{+} + \hat A_c^{-}$) and 
($\hat A_{\bbar}^{+} + \hat A_{\bbar}^{-} + 
\hat A_b^{+} + \hat A_b^{+}$). 
Accordingly, we modify our fitting procedure: instead of
using all eight templates in the fit, we use only four, namely \tm{0},
\tm{1}, \tm{5} and \tm{8}. The values that we obtain for the above
combinations of NP parameters are presented in
Table~\ref{tab:10K_zetasq}. Once again, Re($X^V_{LL}$) proves to be
the weakest link. For the other combinations of NP parameters, the
values extracted from the fit agree with their input values within
$\pm 3.2\sigma$. However, the key point is this: in each test case, a
parameter combination whose input value is nonzero is found from the
fit to be nonzero to at least $7\sigma$.  So, for the NP scenario
\ex{1}, although one cannot draw any conclusions about NP from a
visual examination of the normalized $d\sigma/d\zeta^2_{b \bar b}$
distribution, a fit provides statistically-significant evidence that 
NP is present.

% % % % % % % % % % % % % % % % % % % % % % % % % % % % % % % % % % % 
% % % % % % % % % % % % % % % % % % % % % % % % % % % % % % % % % % % 
% % % % % % % % % % % % % % % % % % % % % % % % % % % % % % % % % % % 
% Table VII
% Fit for zeta^2_bbbar
% initial state - pp
% crossed-diagrams - included
% 10^4 events
{
\renewcommand{\arraystretch}{1.5}
\renewcommand{\tabcolsep}{0.2cm}
\begin{longtable}{|c|c|c|c|c|}
\caption{Values of the combinations of NP parameters extracted from
$d\sigma/d\zeta^2_{b \bar b}$. The integrated luminosity corresponds to 
$10^4$ SM events.} \\ 
\hline 
Test Case & 
Parameter & 
Input Value & 
Fit Result & 
$\chi^2/{\rm d.o.f.}$ \\ 
\hline \hline \endfirsthead

\multicolumn{3}{c}
{{\tablename\ \thetable{} -- continued }} \\
\hline
Test Case &
Parameter & 
Input Value & 
Fit Result & 
$\chi^2/{\rm d.o.f.}$ \\ 
\hline
\endhead

\hline 
\endfoot

\hline 
\endlastfoot

%#####################################################################
%Fit for EX-1
%#####################################################################
%---------------------------------------------------------------------
\texttt{EX-1} & 
SM coefficient &
1 &  
1.2 $\pm$ 0.2 &
1.31 \\ 
%---------------------------------------------------------------------
 & 
$\hat A_{c}^{+}$ + $\hat A_{c}^{-}$ &
64 &
72 $\pm$ 6 &
 \\
%---------------------------------------------------------------------
 & 
$\hat A_{\bar b}^{+}$ + $\hat A_{\bar b}^{-}$ +
$\hat A_{b}^{+}$ + $\hat A_{b}^{-}$ &
32 &
16 $\pm$ 5 &
 \\
%---------------------------------------------------------------------
 &
Re($X^V_{LL}$) &
0 &
3 $\pm$ 1 &
 \\
%---------------------------------------------------------------------
\hline\hline
%#####################################################################
%#####################################################################
%
%
%#####################################################################
%Fit for EX-2
%#####################################################################
%---------------------------------------------------------------------
\texttt{EX-2} & 
SM coefficient &
1 &
0.9 $\pm$ 0.1 &
1.24 \\
%---------------------------------------------------------------------
 & 
$\hat A_{c}^{+}$ + $\hat A_{c}^{-}$ &
0 &
$-$1 $\pm$ 4 &
 \\
%---------------------------------------------------------------------
 & 
$\hat A_{\bar b}^{+}$ + $\hat A_{\bar b}^{-}$ +
$\hat A_{b}^{+}$ + $\hat A_{b}^{-}$ &
25 &
28 $\pm$ 3 &
 \\
%---------------------------------------------------------------------
 &
Re($X^V_{LL}$) &
0 &
0.0 $\pm$ 0.7 &
 \\
%---------------------------------------------------------------------
\hline\hline
%#####################################################################
%#####################################################################
%
%
%#####################################################################
%Fit for EX-3
%#####################################################################
%---------------------------------------------------------------------
\texttt{EX-3} & 
SM coefficient &
1 &
1.3 $\pm$ 0.1 &
1.32 \\ 
%---------------------------------------------------------------------
 & 
$\hat A_{c}^{+}$ + $\hat A_{c}^{-}$ &
0 &
$-$10 $\pm$ 4 &
 \\
%---------------------------------------------------------------------
 & 
$\hat A_{\bar b}^{+}$ + $\hat A_{\bar b}^{-}$ +
$\hat A_{b}^{+}$ + $\hat A_{b}^{-}$ &
25 &
22 $\pm$ 3 &
 \\
%---------------------------------------------------------------------
 &
Re($X^V_{LL}$) &
0 &
$-$1.0 $\pm$ 0.7 &
 \\
%---------------------------------------------------------------------
\hline\hline
%#####################################################################
%#####################################################################
%
%
%#####################################################################
%Fit for EX-4
%#####################################################################
%---------------------------------------------------------------------
\texttt{EX-4} & 
SM coefficient &
1 &
1.0 $\pm$ 0.1 &
0.96 \\ 
%---------------------------------------------------------------------
 & 
$\hat A_{c}^{+}$ + $\hat A_{c}^{-}$ &
0 &
$-$2 $\pm$ 5 &
                                                                \\
%---------------------------------------------------------------------
 & 
$\hat A_{\bar b}^{+}$ + $\hat A_{\bar b}^{-}$ +
$\hat A_{b}^{+}$ + $\hat A_{b}^{-}$ &
61 &
63 $\pm$ 4 &
 \\
%---------------------------------------------------------------------
 &
Re($X^V_{LL}$) &
3 &
2.3 $\pm$ 0.8 &
%---------------------------------------------------------------------
%#####################################################################
%#####################################################################

\label{tab:10K_zetasq}
\end{longtable}
}
% % % % % % % % % % % % % % % % % % % % % % % % % % % % % % % % % % % 
% % % % % % % % % % % % % % % % % % % % % % % % % % % % % % % % % % % 
% % % % % % % % % % % % % % % % % % % % % % % % % % % % % % % % % % % 

Similar fits can be performed with the $d\sigma/d\zeta^2_{b c}$ and
$d\sigma/d\zeta^2_{\bbar c}$ distributions to extract other
combinations of $\hat A_i^{\sigma}$'s. If all the
$d\sigma/d\zeta^2_{12}$'s are combined in one fit, it is possible to
obtain the combinations ($\hat A_{\bbar}^{+} + \hat A_{\bbar}^{-}$),
($\hat A_b^{+} + \hat A_b^{-}$) and ($\hat A_c^{+} + \hat A_c^{-}$).

Analogous to the above example, one may wish to attempt the fit of a
single angular correlation. This, however, is a more complicated task.
As argued earlier, the fits are not sensitive to the individual $\hat
A_i^{\sigma}$'s, but rather to the different kinematic structures that
are present. In the case of the angular correlations [see
  Eq.~(\ref{eq:dsigdcosdcos})], there are two kinematic structures: a
constant term and a term proportional to
$\cos\theta_3^*\cos\theta_\ell^*$.  However, the coefficients of these
pieces involve both SM and NP parameters. Therefore only these
combinations of SM and NP parameters can be extracted.  Furthermore,
note that, in the case of the $\zeta^2_{12}$ distributions, each
template is proportional to a single kinematic structure. For example,
for $d\sigma/d\zeta^2_{b\bbar}$, \tm{5} is only sensitive to $h_c^{b
  \bbar}$. On the other hand, in the case of the angular correlations,
each template contains both the constant piece and the
$\cos\theta_3^*\cos\theta_\ell^*$ piece.

In order to work around these difficulties, we proceed as
follows. First, we fix the weight of the SM contribution\footnote{Note
  that this could also have been done for the fits to the
  $\zeta^2_{12}$ distributions. However, in the case of an angular
  correlation it {\it must} be done.} to be 1.0. Second, the templates
themselves have to be reorganized. For example, consider
$d\sigma/d\cos\theta_c^*d\cos\theta_{\ell}^*$.  Here a template
defined as (\tm{5} + \tm{6}) would be proportional to the constant
piece, and another defined as (\tm{5} $-$ \tm{6}) would be
proportional to the $\cos\theta_c^*\cos\theta_{\ell}^*$
piece\footnote{This holds as long as the values of $\hat A_{c}^{+}$
  and $\hat A_{c}^{-}$ used to generate \tm{5} and \tm{6},
  respectively, are identical.}.  The coefficients of these two
modified templates would then be expected to yield the values of $(
\hat A_{c}^{+} + \hat A_{c}^{-} + \hat A_{\bbar}^{+} + \hat
A_{\bbar}^{-} + \hat A_{b}^{+} + \hat A_{b}^{-} )$ and $( \hat
A_{c}^{+} - \hat A_{c}^{-} - \frac{1}{3} \{ \hat A_{\bbar}^{+} - \hat
A_{\bbar}^{-} + \hat A_{b}^{+} - \hat A_{b}^{-} \} )$.

The results of the fit for the different \ex{i} are presented in
Table~\ref{tab:10K_angcorr}. For all four \ex{i}, the agreement
between best-fit and input values is very good for $( \hat A_{c}^{+} +
\hat A_{c}^{-} + \hat A_{\bbar}^{+} + \hat A_{\bbar}^{-} + \hat
A_{b}^{+} + \hat A_{b}^{-} )$. The key point is that, in all cases,
this combination of NP parameters is definitely nonzero. For $( \hat
A_{c}^{+} - \hat A_{c}^{-} - \frac{1}{3} \{ \hat A_{\bbar}^{+} - \hat
A_{\bbar}^{-} + \hat A_{b}^{+} - \hat A_{b}^{-} \} )$ the error bars
are larger: the best-fit and input values agree to within
1-3$\sigma$.  Nevertheless, a fit to a single angular correlation
can provide statistically-significant evidence that NP is present. The
measurement of an angular correlation would, however, most likely be
more challenging than the measurement of a $\zeta^2_{12}$
distribution, which is essentially an invariant mass-squared
distribution. Hence it is very likely that NP, if present, will be
discovered first in a $\zeta^2_{12}$ distribution.

% % % % % % % % % % % % % % % % % % % % % % % % % % % % % % % % % % % 
% % % % % % % % % % % % % % % % % % % % % % % % % % % % % % % % % % % 
% % % % % % % % % % % % % % % % % % % % % % % % % % % % % % % % % % % 
% Table VIII
% Fit for costheta_c*costheta_l
% initial state - pp
% crossed-diagrams - included
% 10^4 events
{
\renewcommand{\arraystretch}{1.5}
\renewcommand{\tabcolsep}{0.2cm}
\begin{longtable}{|c|c|c|c|c|}
\caption{Values of the combinations of NP parameters extracted from
$d\sigma/d\cos\theta_c^* d\cos\theta_{\ell}^*$. The integrated luminosity
corresponds to $10^4$ SM events. The weight of the SM contribution is
fixed to be 1.0.} \\ 
\hline 
Test Case & 
Parameter &
Input Value &
Fit Result &
$\chi^2/{\rm d.o.f.}$ \\ 
\hline \hline \endfirsthead

\multicolumn{3}{c}
{{\tablename\ \thetable{} -- continued }} \\
\hline
Test Case &
Parameter &
Input Value &
Fit Result &
$\chi^2/{\rm d.o.f.}$ \\
\hline
\endhead

\hline 
\endfoot

\hline 
\endlastfoot

%#####################################################################
%Fit for EX-1
%#####################################################################
%---------------------------------------------------------------------
\texttt{EX-1} & 
$\hat A_{c}^{+} + \hat A_{c}^{-} +
\hat A_{\bar b}^{+} + \hat A_{\bar b}^{-} +
\hat A_{b}^{+} + \hat A_{b}^{-}$ &
96 &
97 $\pm$ 1 &
1.19 \\
%---------------------------------------------------------------------
 & 
$\hat A_{c}^{+} - \hat A_{c}^{-} - 
\frac{1}{3}
\big( 
\hat A_{\bar b}^{+} - \hat A_{\bar b}^{-} +
\hat A_{b}^{+} - \hat A_{b}^{-}
\big)$ &
0 &
26 $\pm$ 11 &
 \\
%---------------------------------------------------------------------
\hline\hline
%#####################################################################
%#####################################################################
%
%
%#####################################################################
%Fit for EX-2
%#####################################################################
%---------------------------------------------------------------------
\texttt{EX-2} & 
$\hat A_{c}^{+} + \hat A_{c}^{-} +
\hat A_{\bar b}^{+} + \hat A_{\bar b}^{-} +
\hat A_{b}^{+} + \hat A_{b}^{-}$ &
25 &
26 $\pm$ 1 &
1.00 \\
%---------------------------------------------------------------------
 & 
$\hat A_{c}^{+} - \hat A_{c}^{-} - 
\frac{1}{3}
\big( 
\hat A_{\bar b}^{+} - \hat A_{\bar b}^{-} +
\hat A_{b}^{+} - \hat A_{b}^{-}
\big)$ &
$-$8.33 &
$-$10 $\pm$ 7 &
 \\
%---------------------------------------------------------------------
\hline\hline
%#####################################################################
%#####################################################################
%
%
%#####################################################################
%Fit for EX-3
%#####################################################################
%---------------------------------------------------------------------
\texttt{EX-3} & 
$\hat A_{c}^{+} + \hat A_{c}^{-} +
\hat A_{\bar b}^{+} + \hat A_{\bar b}^{-} +
\hat A_{b}^{+} + \hat A_{b}^{-}$ &
25 &
26 $\pm$ 1 &
1.01 \\
%---------------------------------------------------------------------
 & 
$\hat A_{c}^{+} - \hat A_{c}^{-} - 
\frac{1}{3}
\big( 
\hat A_{\bar b}^{+} - \hat A_{\bar b}^{-} +
\hat A_{b}^{+} - \hat A_{b}^{-}
\big)$ &
2.33 &
$-$3 $\pm$ 7 &
 \\
%---------------------------------------------------------------------
\hline\hline
%#####################################################################
%#####################################################################
%
%
%#####################################################################
%Fit for EX-4
%#####################################################################
%---------------------------------------------------------------------
\texttt{EX-4} & 
$\hat A_{c}^{+} + \hat A_{c}^{-} +
\hat A_{\bar b}^{+} + \hat A_{\bar b}^{-} +
\hat A_{b}^{+} + \hat A_{b}^{-}$ &
61 &
64 $\pm$ 1 &
0.97 \\
%---------------------------------------------------------------------
 & 
$\hat A_{c}^{+} - \hat A_{c}^{-} - 
\frac{1}{3}
\big( 
\hat A_{\bar b}^{+} - \hat A_{\bar b}^{-} +
\hat A_{b}^{+} - \hat A_{b}^{-}
\big)$ &
$-$20.33 &
$-$31$\pm$ 9 &
%---------------------------------------------------------------------
%#####################################################################
%#####################################################################

\label{tab:10K_angcorr}
\end{longtable}
}

The simplest approach towards the fitting of the angular correlations
would have been to fit them to the functional form 
$a_1 + a_2\cos\theta_3^*\cos\theta_\ell^*$, as is the usual procedure
for measuring $\kappa_{t \tbar}$. However, this possibility is 
precluded due to the fact that the event-selection criteria described 
in \textit{Fit 2} distorts the shape of the correlation.
We have used a somewhat simple-minded approach to deal with the
identical $\bbar$'s in the final state. It is certainly possible that
experimentalists will find a better way to deal with this situation
(perhaps through the use of some sophisticated multivariate technique,
such as neural networks or boosted decision trees) and that such an
approach would lead to less distortion of the shape of the 
correlation.

Finally, we consider the full fit involving all six observables, with
statistics corresponding to 10$^4$ events for the SM. The results are
presented in Table~\ref{tab:10K_full}. Apart from Re($X^V_{LL}$) in
\ex{1}, the values of all NP parameters agree with their input values
within $\pm 1.7\sigma$.

% % % % % % % % % % % % % % % % % % % % % % % % % % % % % % % % % % % 
% % % % % % % % % % % % % % % % % % % % % % % % % % % % % % % % % % % 
% % % % % % % % % % % % % % % % % % % % % % % % % % % % % % % % % % % 
% Table IX
% Fit for all observables
% initial state - pp
% crossed-diagrams - included
% 10^4 events
{
\renewcommand{\arraystretch}{1.5}
\renewcommand{\tabcolsep}{0.2cm}
\begin{longtable}{|c|l@{}l|c|}
\caption{Values of the NP parameters extracted from a fit to all six
observables. The integrated luminosity corresponds to $10^4$ SM events.}
\\ 
\hline 
Test Case & 
\multicolumn{2}{c|}{Fit Results} & 
$\chi^2/{\rm d.o.f.}$ \\ 
\hline \hline \endfirsthead

\multicolumn{4}{c}
{{\tablename\ \thetable{} -- continued }} \\
\hline
Test Case &
\multicolumn{2}{c|}{Fit Results} &
$\chi^2/{\rm d.o.f.}$ \\
\hline
\endhead

\hline 
\endfoot

\hline
\endlastfoot

%#####################################################################
%Fit for EX-1
%#####################################################################
%---------------------------------------------------------------------
\texttt{EX-1} & 
\multicolumn{2}{l|}{SM coefficient = 1.00 $\pm$ 0.01} &
1.18 \\
%---------------------------------------------------------------------
 & 
$\hat A_{\bar b}^{+}$ = 36 $\pm$ 9 &  
$\hat A_{\bar b}^{-}$ = 28 $\pm$ 9 &
 \\
%---------------------------------------------------------------------
 &
$\hat A_{b}^{+}$ = $-$22 $\pm$ 8 &
$\hat A_{b}^{-}$ = $-$11 $\pm$ 8 &
 \\
%---------------------------------------------------------------------
 &
$\hat A_{c}^{+}$ = 47 $\pm$ 9 &
$\hat A_{c}^{-}$ = 18 $\pm$ 9 &
 \\
%---------------------------------------------------------------------
 &
\multicolumn{2}{l|}{Re($X^V_{LL}$) = 0.31 $\pm$ 0.09} &
 \\
%---------------------------------------------------------------------
\hline\hline
%#####################################################################
%#####################################################################
%
%
%#####################################################################
%Fit for EX-2
%#####################################################################
%---------------------------------------------------------------------
\texttt{EX-2} & 
SM coefficient = 0.988 $\pm$ 0.009 &
 &
0.92 \\ 
%---------------------------------------------------------------------
 & 
$\hat A_{\bar b}^{+}$ = 6 $\pm$ 6 &  
$\hat A_{\bar b}^{-}$ = $-$3 $\pm$ 6 &
 \\
%---------------------------------------------------------------------
 &
$\hat A_{b}^{+}$ = 23 $\pm$ 5 &
$\hat A_{b}^{-}$ = 1  $\pm$ 6 &
 \\
%---------------------------------------------------------------------
 &
$\hat A_{c}^{+}$ = 0 $\pm$ 6 &
$\hat A_{c}^{-}$ = $-$1 $\pm$ 6 &
 \\
%---------------------------------------------------------------------
 &
\multicolumn{2}{l|}{Re($X^V_{LL}$) = 0.03 $\pm$ 0.05} &
 \\
%---------------------------------------------------------------------
\hline\hline
%#####################################################################
%#####################################################################
%
%
%#####################################################################
%Fit for EX-3
%#####################################################################
%---------------------------------------------------------------------
\texttt{EX-3} & 
SM coefficient = 1.013 $\pm$ 0.009 &
 &
0.94 \\
%---------------------------------------------------------------------
 & 
$\hat A_{\bar b}^{+}$ = 4 $\pm$ 6 &  
$\hat A_{\bar b}^{-}$ = $-$3 $\pm$ 6 &
 \\
%---------------------------------------------------------------------
 &
$\hat A_{b}^{+}$ = 16 $\pm$ 5 &
$\hat A_{b}^{-}$ = 8 $\pm$ 6 &
 \\
%---------------------------------------------------------------------
 &
$\hat A_{c}^{+}$ = 1 $\pm$ 6 &
$\hat A_{c}^{-}$ = $-$2 $\pm$ 6 &
 \\
%---------------------------------------------------------------------
 &
\multicolumn{2}{l|}{Re($X^V_{LL}$) = 0.03 $\pm$ 0.05} &
 \\
%---------------------------------------------------------------------
\hline\hline
%#####################################################################
%#####################################################################
%
%
%#####################################################################
%Fit for EX-4
%#####################################################################
%---------------------------------------------------------------------
\texttt{EX-4} & 
SM coefficient = 1.00 $\pm$ 0.01 &
 &
0.99 \\
%---------------------------------------------------------------------
 & 
$\hat A_{\bar b}^{+}$ = 47 $\pm$ 7 &  
$\hat A_{\bar b}^{-}$ = $-$12 $\pm$ 7 &
 \\
%---------------------------------------------------------------------  
 &
$\hat A_{b}^{+}$ = 34 $\pm$ 7 &
$\hat A_{b}^{-}$ = $-$9 $\pm$ 7 &
 \\
%---------------------------------------------------------------------
 &
$\hat A_{c}^{+}$ = 8 $\pm$ 8 &
$\hat A_{c}^{-}$ = $-$8 $\pm$ 8 &
 \\
%---------------------------------------------------------------------
                                                                &
\multicolumn{2}{l|}{Re($X^V_{LL}$) = 2.91 $\pm$ 0.05}           &
%---------------------------------------------------------------------
%#####################################################################
%#####################################################################

\label{tab:10K_full}
\end{longtable}
}
% % % % % % % % % % % % % % % % % % % % % % % % % % % % % % % % % % % 
% % % % % % % % % % % % % % % % % % % % % % % % % % % % % % % % % % % 
% % % % % % % % % % % % % % % % % % % % % % % % % % % % % % % % % % % 

To sum up, the above simulations demonstrate that, even in the short
term, it is possible to detect the presence of NP in top decay through
the measurement of the invariant mass-squared distributions and/or the
angular correlations. This can be done by comparing the
measured shapes of the distributions/correlations with the SM
predictions. More sensitivity can be obtained by performing fits to
extract combinations of NP parameters. If all six distributions and
correlations can be measured, a combined fit can be performed to
extract all the NP parameters. The determination of which parameters
are nonzero allows for a partial identification of the NP.

\subsection{Long term}

As noted above, the effective cross section in $\ggprocess$ is $\sim
0.1$ pb. The LHC is projected to deliver 3000 fb$^{-1}$ worth
of data by the year 2030 \cite{LHCprojection}. Assuming this
integrated luminosity and a $b$-tagging efficiency of 70\% for each of
the three $b$ or $\bbar$'s in the final state, one obtains $\approx
10^5$ events of the type $p p \to t \tbar \to (b \bbar c) (\bbar
\ell^- \bar \nu_{\ell})$ from the SM. This is the number of events in
our long-term simulations.

By 2030, all six distributions and correlations will, in all
likelihood, have been measured.  For this reason we consider only the
fit to all distributions/correlations with $10^5$ SM events. The
corresponding results have already been presented in
Table~\ref{tab:fit3}.  Apart from Re($X^V_{LL}$) in \ex{1}, the
best-fit values of all NP parameters differ from their input values by
at most $1\sigma$. The errors on the $\hat{A}_i^\sigma$'s are
typically in the range 1.75-2.75.  Thus, any $\hat{A}_i^\sigma$ that
is $\gsim 10$ will be found to be nonzero at a
statistically-significant level. In this way it will be possible to
determine which NP parameters are nonzero, thus producing an
identification of the NP.

\section{Conclusions}

In Ref.~\cite{companion}, the companion paper, new physics (NP) in the
decay $\tbbc$ is considered. There, ten dimension-6 NP operators
contributing to $\tbbc$ are delineated, and two types of observables
are identified that can be used to search for this NP in the process
$\ggprocess$. They are (i) invariant mass-squared distributions
involving the $\{b,c\}$, $\{\bbar,c\}$, or $\{b,\bbar\}$ quark pairs
coming from $\tbbc$, and (ii) angular correlations between the
$\ell^-$ coming from the $\tbar$ decay and one of $\bbar$, $b$ or $c$
coming from the $t$ decay. It is further shown that the NP
contributions to these observables can be written in terms of certain
combinations of the NP couplings, denoted as $\Ahat$. In the present
paper we examine the prospects for detecting and identifying such NP
at the LHC.

The first step is to develop an algorithm to extract the $\Ahat$'s and
Re$\left(X^V_{LL}\right)$ from the observables. From the analytical
expressions obtained in Ref.~\cite{companion} [summarized here in
  Eqs.~(\ref{eq:dsigdM}) and (\ref{eq:dsigdcosdcos})], we learn that
the NP contribution to the observables can be represented as a linear
combination of pieces proportional to the different $\Ahat$'s and
Re$\left(X^V_{LL}\right)$. Using this idea, we perform a Monte-Carlo
simulation using \madg to compute ``templates,'' which are the
contributions of the SM, each $\Ahat$ and Re$\left(X^V_{LL}\right)$ to
the observables.  We also generate Monte-Carlo data for four possible
NP scenarios.  For each of these scenarios, we extract the NP
parameters simply by obtaining the weights with which the templates
must be combined to reproduce the Monte-Carlo data.

Although the fit algorithm is based on a simple premise, there are two
issues that must be taken into account.  First, the construction of
the observables requires distinguishing the decay products of the $t$
from those of the $\bar t$. However, the final state contains two
$\bar b$'s, which are indistinguishable, at least in some parts of
phase space. We designate the $\bar b$ that yields the smaller value
of $|m_t - \sqrt{(p_{\bar b} + p_b + p_c)^2}|$ as that having come
from the $t$ decay. However, if both $\bar b$'s in the event yield
sufficiently small values of this quantity (less than 15$\Gamma_t$),
then we consider them to be indistinguishable and exclude such events
from the analysis. Second, the contribution to $t \tbar$ production
from a $q \bar q$ initial state is not included in the analytical
expressions. However, this must be taken into account as there are no
known algorithms that can efficiently separate $t \tbar$ pairs coming
from gluon fusion from those occuring due to $q \bar q$ annihilation.

In order to examine the prospects for detecting the presence of NP in
$\tbbc$, and for its identification, we perform further
simulations of the distributions/correlations.  The simulations are
done for either short-term or long-term measurements at the LHC. For
the short-term analysis we use $10^4$ events of the type $p p \to t
\tbar \to (b \bbar c) (\bbar \ell^- \bar \nu_{\ell})$. This is
expected to be delivered by 2020-2021. For the long term we use $10^5$
events, which is projected by the year 2030.

In the short term not all distributions/correlations may be measured,
and what can be learned about the NP depends on what measurements have
been made.  In the presence of a sufficiently-large NP contribution to
$\tbbc$, the shapes of the distributions/correlations can be
significantly modified.  Thus, NP in $\tbbc$ may be inferred by
observing a clear difference between the shape of a measured
distribution and its SM prediction.  Even if there is no discernible
difference in the shapes, it may still be possible to obtain
information about NP contributions.  Using the above algorithm with a
slight modification, one can perform a fit to a single
distribution. In this case, not all the individual $\Ahat$'s are
extracted, but rather certain combinations of the $\Ahat$'s.  We show
that, even for a scenario in which the presence of NP does not induce
a substantial change in the shape of the distribution, a fit may still
yield statistically-significant evidence that NP is present. Finally,
if all six distributions/correlations are measured, we can use the
algorithm to perform a simultaneous fit on all the observables to
extract Re$\left(X^V_{LL}\right)$ and all the $\Ahat$'s separately.
In the examples studied, we find that the values of all NP parameters
agree with their input values within $\pm 1.7\sigma$. Although the
errors are large, this provides an approximate determination of the
values of the NP parameters. More importantly it allows us to infer
that a non-zero NP contribution to $\tbbc$ exists.

In the long term, it is likely that all six distributions/correlations
will be measured. Furthermore, the availability of larger statistics
will lead to an improvement in the quality of the fits. We find that,
with $10^5$ events, the best-fit values of all NP parameters differ
from their input values by at most $1\sigma$.  Thus, if NP is present
in $\tbbc$, the fit will allow the determination of its nature.

\bigskip
\noindent
{\bf Acknowledgments}: The authors wish to thank the MadGraph and
FeynRules Teams for extensive discussions about MadGraph and
FeynRules, respectively.  PS would like to thank Georges Azuelos
for helpful discussions about MadGraph. The authors also wish to thank
S. Judge and J. Melendez for collaboration at an early stage of this
work and J. Kiers and R. Rezvani for helpful discussions. This work 
was financially supported by NSERC of Canada (DL, PS).  This work has
been partially supported by ANPCyT under grant No. PICT-PRH 2009-0054
and by CONICET (AS).  The work of KK was supported by the U.S.\ 
National Science Foundation under Grant PHY--1215785.  KK also
acknowledges sabbatical support from Taylor University.

%%%%%%%%%%%%%%%%%%%% REFERENCES %%%%%%%%%%%%%%%%%%%%%%%%%%%%%%%%%%%%%%

\end{document}